%% file: quasinormalmodes.tex
\documentclass[prd,twocolumn,nofootinbib,superscriptaddress,amsmath,amssymb,groupedaddress]{revtex4-1}
\include{preamble}

\begin{document}

\title{Probing beyond-Kerr spacetimes with inspiral-ringdown corrections to gravitational waves}

\author{Zack Carson}
\author{Kent Yagi}

\affiliation{Department of Physics, University of Virginia, Charlottesville, Virginia 22904, USA}

\date{\today}


\begin{abstract}
Gravitational waves from the explosive merger of distant black holes are encoded with details regarding the complex extreme-gravity spacetime present at their source.
Famously described by the Kerr spacetime metric for rotating black holes in general relativity, what if effects beyond this theory are present?
One way to efficiently test this hypothesis is to first obtain a metric which parametrically deviates from the Kerr metric in a model-independent way.
Given such a metric, one can then predict the ensuing corrections to both the inspiral and ringdown portions of the gravitational waveform for black holes present in the new spacetime.
With these tools in hand, one can then test gravitational wave signals for such effects by two different methods, (i) inspiral-merger-ringdown consistency test, and (ii) parameterized test.
In this paper, we demonstrate the exact recipe one needs to do just this.
We first derive parameterized corrections to the waveform inspiral, ringdown, and remnant properties for a generic non-Kerr spacetime and apply this to two example beyond-Kerr spacetimes each parameterized by a single non-Kerr parameter.
We then predict the beyond-Kerr parameter magnitudes required in an observed gravitational wave signal to be statistically inconsistent with the Kerr case in general relativity.
We find that the two methods give very similar bounds. 
The constraints found with existing gravitational-wave events are comparable to those from x-ray observations, while future gravitational-wave observations using Cosmic Explorer (Laser Interferometer Space Antenna) can improve such bounds by two (three) orders of magnitude.

\end{abstract}

\maketitle


\section{Introduction}\label{sec:intro}
The famous observation of gravitational waves (GWs) from the distant coalescence of two black holes (BHs) has opened the window for a new probe of the \emph{extreme gravity} spacetimes~\cite{Abbott_IMRcon2,Yunes_ModifiedPhysics,Berti_ModifiedReviewSmall} present around such objects.
Observed by the Laser Interferometer Gravitational-wave Observatory~\cite{aLIGO} (LIGO) on September 14, 2015, the impressive observation of GW150914~\cite{GW150914} was only the first of eleven GW events detected~\cite{GW_Catalogue}.
Encoded within the GWs emanating from such extreme events across the universe is a treasure-trove of information regarding the local spacetime present around the BHs.
This incredible opportunity allows us for the first time to test the theory of gravity describing these extreme-gravity environments, where the fields are highly strong, non-linear, and dynamical.

The currently accepted model of gravity is described by the theory of general relativity (GR) prescribed by Einstein over a century ago.
Just as previously, Newtonian's prescription of gravity was relentlessly tested, we now must put GR to the test.
Since then, GR has been subject to observations in a variety of different spacetime environments, including within the solar system where gravity is weak and static~\cite{Will_SolarSystemTest}, in strong and static-field binary pulsar systems~\cite{Stairs_BinaryPulsarTest,Wex_BinaryPulsarTest}, in large-scale cosmological observations~\cite{Ferreira_CosmologyTest,Clifton_CosmologyTest,Joyce_CosmologyTest,Koyama_CosmologyTest,Salvatelli_CosmologyTest}, and finally in the extreme-gravity observations of GWs~\cite{Abbott_IMRcon2,Yunes_ModifiedPhysics,Berti_ModifiedReviewSmall,TheLIGOScientific:2016pea,Abbott_IMRcon,Monitor:2017mdv,Abbott:2018lct}.
In every scenario, all such tests of GR have been found to agree remarkably with Einstein's theory.
With several future GW detector improvements~\cite{Ap_Voyager_CE} and third-generation detectors~\cite{Ap_Voyager_CE,ET,LISA,B-DECIGO,DECIGO,TianQin} planned, there is a hope that, if indeed small deviations beyond GR are present in nature, enhanced detector sensitivities will allow us to observe them.

Why must we continue to test GR, even with its unprecedented successes in the last century?
While GR can accurately explain many of our observations of the universe, there remain several open questions still debated upon today.
Prominent among the possible solutions are new, alternative theories of gravity that go beyond GR.
In fact, the scenario in which a modified theory of gravity is activated in extreme-gravity spacetimes, while reducing to GR in the weak-gravity environments where a majority of our tests have been performed in, is a possibility.
In particular, several alternative theories of gravity have been introduced which explain some of these open questions, such as: the theory of inflation in the early universe~\cite{Joyce:2014kja,Clifton:2011jh,Famaey:2011kh,Koyama:2015vza}; ``dark energy's'' impact on the universe's accelerated expansion~\cite{Jain:2010ka,Salvatelli:2016mgy,Koyama:2015vza,Joyce:2014kja}; the universe's present matter/anti-matter asymmetry~\cite{Clifton:2011jh,Famaey:2011kh}; the unification of quantum mechanics with GR~\cite{Clifton:2011jh,Joyce:2014kja,Famaey:2011kh,Milgrom:2008rv,Jain:2010ka,Koyama:2015vza}; and finally ``dark matter's'' influence on the inconsistent rotation curves of galaxies throughout the universe~\cite{Famaey:2011kh,Milgrom:DarkMatter,Milgrom:2008rv,Clifton:2011jh,Joyce:2014kja}.
In order to put answers to these questions, we must continue to test GR in the most exotic spacetimes possible, with the final hope of detecting possible deviations.

To test the current theory of gravity, one needs to first develop a new spacetime metric as a solution to modified field equations.
In GR, this spacetime metric is described by the Kerr result $g_{\alpha\beta}^\K$ for rotating BHs.
In order to perform the tests in an efficient, theory-agnostic way in beyond GR, one would presumably introduce parameterized deviations from the Kerr metric which, when vanishing, reproduces the Kerr result again.
In this scenario, one or more non-GR deviation parameters could be observationally constrained in a model-independent way that requires no prior theoretical knowledge.
One could then map such bounds to those on theoretical parameters in specific non-GR theories.
To date, several such metrics have been developed~\cite{Collins:2004ex,Glampedakis:2005cf,Vigeland:2011ji,PhysRevD.81.024030,Johannsen:2011dh,Manko,Konoplya:2020hyk}, each of which are stationary, axisymmetric, asymptotically flat, and contain one or more parameters deviating from the Kerr metric. 
For example, Johannsen developed a more general parameterized metric with separable geodesic equations in~\cite{Johannsen:2015pca}, followed up by an even more general metric preserving the same symmetries found in~\cite{Papadopoulos:2018nvd,Carson_BumpyPhotonRings}.
Several of the above parameterized metrics can then be mapped to many known BH solutions found in the literature~\cite{Papadopoulos:2018nvd,Randall:1999ee,Aliev:2005bi,Jai-akson:2017ldo,Ding:2019mal,Kerr-Sen,Kanti_EdGB,Maeda:2009uy,Sotiriou:2014pfa,Ayzenberg:2014aka,Jackiw:2003pm,Yagi_dCS,Yunes_dcs,Bardeen,Kumar:2019uwi,Pani:2011gy,Kumar:2020hgm}, with popular transformations for the latter two metrics tabulated in~\cite{Johannsen:2015mdd,Carson_BumpyPhotonRings}.

In this paper, we focus our attentions on two parameterized metrics and see how well one can probe such spacetimes with current and future GW observations. The first one was derived by Johannsen and Psaltis (JP)~\cite{Johannsen:2011dh}, which has a single deviation parameter $\epsilon_3$. This spacetime is an example of the more general metric found in~\cite{Johannsen:2015pca}. 
The second one is motivated by Johannsen in~\cite{Johannsen:2015pca}, where a new deviation parameter $\beta$\footnote{$\beta$ is introduced in $\Delta = r^2 - 2 M r + a^2$ as $\Delta \to \bar \Delta = \Delta + \beta$, where $M$ and $a$ characterize the mass and spin of a BH while $r$ is the radial coordinate.} was introduced into the more general metric in~\cite{Johannsen:2015pca}.
Here, we remove all non-Kerr deviation parameters with the exception of $\beta$ to form the singly-parameterized ``modified-$\Delta$'' (mod.-$\Delta$, or ``MD'' in superscripts/subscripts) metric.
Because these single-parameter spacetimes have been obtained from the more general beyond-GR metrics which can be mapped to several known BH solutions, they make ideal candidates for testing GR in a simple way.

With a model-independent beyond-GR metric in hand, one next needs to find the modifications to the gravitational waveform imparted under the new spacetime.
As accomplished in similar works by the same authors~\cite{Carson_QNMPRD,Carson_QNM_PRL} in the Einstein-dilaton Gauss-Bonnet theory of gravity~\cite{Kanti_EdGB,Maeda:2009uy,Sotiriou:2013qea,Yagi:2015oca}, one can obtain analytic expressions for various corrections to the gravitational waveform in an alternative theory of gravity.
In this paper, given an arbitrary spacetime metric $g_{\alpha\beta}^\X$, we show how one can obtain corrections to the GW inspiral, ringdown quasinormal modes (QNMs), and the remnant BH's mass and spin.
When inserted into the standard GR gravitational waveform, these singly-parameterized corrections can be used to test future incoming signals for deviations from GR.
See also Ref.~\cite{Suvorov:2019qow} where it was detailed how one can test beyond-GR theories of gravity, even for Kerr BHs~\cite{Psaltis:2007cw} if one considers their perturbations.

With a generalized beyond-GR metric and its resulting corrections to the gravitational waveform template, one needs to test the observed GW signals for deviations present within.
Specifically, we focus our attention on the so-called inspiral-merger-ringdown (IMR) consistency tests of GR~\cite{Ghosh_IMRcon,Ghosh_IMRcon2,Abbott_IMRcon,Abbott_IMRcon2}. 
In this application, one tests the consistency between the inspiral and merger-ringdown GW signals to predict the possibility of emergent non-Kerr effects present in the observed signal.
In particular, we estimate (with a Fisher analaysis~\cite{Poisson:Fisher,Berti:Fisher,Yagi:2009zm}) the final BH's mass and spin individually from the inspiral signal, and then from the merger-ringdown signal.
If both predictions show significant disagreement from each other, one can conclude with evidence of non-Kerr effects present in the observed signal (provided systematic errors are under control).

In addition to the IMR consistency tests, we also test the gravitational waveform in a parameterized way.
Indicated as the parameterized tests of GR throughout the following paper, we begin by introducing corrections to the waveform inspiral, merger-ringdown, and to the remnant BH properties.
All such corrections are parameterized by the single JP or mod.-$\Delta$ parameters $\epsilon_3$ and $\beta$ which allows for a convenient test.
We assume the waveform is described by GR ($\epsilon_3=0$ or  $\beta=0$ ) and estimate the resulting root-mean-square uncertainties on the non-GR parameters.
Such variations then describe the ``wiggle room'' such non-GR parameters have to still remain consistent within the GW detector's noise, and can be taken as an upper-bound constraint.

\renewcommand{\arraystretch}{1.2}
\begin{table*}
        \centering
        \begin{tabular}{|cc|c|c|c|c|}
             \cline{1-6}
			 && \multicolumn{2}{c|}{}&\multicolumn{2}{c|}{}\\[-1em]
             && \multicolumn{2}{c|}{$\epsilon_3$ (JP~\cite{Johannsen:2011dh})}  & \multicolumn{2}{c|}{$\beta$ (mod.-$\Delta$~\cite{Johannsen:2015mdd,Johannsen:2015pca})}\\
             && \multicolumn{2}{c|}{}&\multicolumn{2}{c|}{}\\[-1em]
             && IMR & Param. &  IMR & Param.\\
             &&&&&\\[-1em]
            \hline
            &&&&&\\[-1em]
            \multirow{2}{*}{O2~\cite{aLIGO}} & GW150914~\cite{GW150914} & (7)$^*$ & (5)$^*$ & (2)$^*$ & (1)$^*$\\
            \cline{2-2}
            & GW170729~\cite{GW170729} & (10)$^*$ & (14)$^*$ & (14)$^*$ & (11)$^*$ \\
            &&&&&\\[-1em]
            \hline
            &&&&&\\[-1em]
            \multirow{2}{*}{CE~\cite{Ap_Voyager_CE}} & GW150914~\cite{GW150914} & 0.05 & 0.05 & 0.05 & 0.02\\
            \cline{2-2}
            & GW170729~\cite{GW170729} & 0.6 & 0.5 & 0.06 & 0.07\\
            &&&&&\\[-1em]
            \hline
             &&&&&\\[-1em]
            \multirow{2}{*}{CE+LISA~\cite{Ap_Voyager_CE,LISA}} & GW150914~\cite{GW150914} & 0.02 & 0.03 & $5\times10^{-3}$ & $4\times10^{-3}$\\
            \cline{2-2}
            & GW170729~\cite{GW170729} & 0.05 & 0.09 & 0.05 & 0.03\\
            &&&&&\\[-1em]
            \hline
			&&&&&\\[-1em]
            \multirow{2}{*}{LISA~\cite{LISA}} & EMRI & $(2\times10^{-3})^\dagger$ & $10^{-3}$ & ($2\times10^{-4}$)$^\dagger$ & $10^{-4}$ \\
            \cline{2-2}
            & SMBHB & 0.02 & 0.01 & $10^{-3}$ & $10^{-3}$\\
            \hline
        \end{tabular}
        \caption{
        Summary of results obtained in our analysis for both the Johannsen-Psaltis and modified-$\Delta$ metrics. Here we compare constraints on the deviation parameters $\epsilon_3$ and $\beta$ obtained via the inspiral-merger-ringdown consistency tests of GR (IMR), and the parameterized tests of GR (Param.) for each gravitational-wave event and detector considered. In particular, bounds are presented for GW150914-like events ($m_1=36\text{ M}_\odot$, $m_2=29\text{ M}_\odot$), GW170729-like events ($m_1=50.6\text{ M}_\odot$, $m_2=34.4\text{ M}_\odot$), EMRIs ($m_1=10^6\text{ M}_\odot$, $m_2=10\text{ M}_\odot$), and super-massive black hole binaries (SMBHBs,  $m_1=10^6\text{ M}_\odot$, $m_2=5\times10^4\text{ M}_\odot$). Observe that the bounds with the two methods are comparable in all cases presented here.\\
        $*$  \scriptsize{Constraints with the aLIGO O2 detector are not as reliable because they fall beyond the small-deviation approximation made when deriving ppE parameters.}\\
        $\dagger$  \scriptsize{Constraints from EMRIs with IMR consistency tests may not be accurate since the IMRPhenomD waveforms were calibrated to numerical relativity simulations with mass ratios only up to 1:18. In the parameterized test, all such numerical relativity (NR) fits have been removed, and integrations stopped before the merger to avoid such inaccuracies.}
        }\label{tab:results}
\end{table*}

We refer readers to related works on testing beyond-Kerr spacetimes with GWs.
References~\cite{Gair:2011ym,Moore:2017lxy} construct an approximate, multipolar gravitational waveform suitable for extreme-mass-ratio-inspirals (EMRIs) detectable by space-based detector LISA for inspiral using the analytic kludge method from a beyond-Kerr ``bumpy'' spacetime, which can be use to test GR with GW signals by placing constraints on the deviations, as was considered for future LISA observations.
Reference~\cite{Chua:2018yng} considered similar EMRI analytic kludge waveforms and performed a Bayesian model selection analysis for distinguishing Kerr and beyond-Kerr models.
Additionally, Refs.~\cite{Barack:2006pq,Glampedakis:2005cf} considered quadrupole corrections to the GR Kerr analytic kludge waveforms for EMRIs in a bumpy spacetime to consider the accuracy with which LISA could constrain such deformations.
Reference~\cite{Xin:2018urr} considered the JP metric considered in this paper to build a parameterized EMRI waveform and test it with future space-based observations.
Even more recently Ref.~\cite{Cardenas-Avendano:2019zxd} considered a singly-parameterized beyond-Schwarzschild (non-spinning BHs) metric and derive corrections to the inspiral waveform to place constraints on previous LVC detections.
See also Ref.~\cite{Maselli:2019mjd} where similar corrections to the QNMs were made, and constraints with future observations of multiple GW events were quantified.

The analysis presented here differs from above at least in a few ways. 
For example, we not only consider different beyond-Kerr spacetimes than the ones considered above (except for~\cite{Xin:2018urr}), but we additionally find corrections to the ringdown waveform and also to the remnant BH properties, all up to quadratic order in BH rotation.
Additionally, while all of the above analyses focus on waveforms suitable for EMRIs detectable by LISA, in our analysis, we find corrections to the commonly-used IMRPhenomD gravitational waveform which is more suited to comparable-mass systems (this waveform has been calibrated for mass-ratios up to $1:18$, significantly smaller than that for EMRIs of $\sim 1:10^5$).

In this paper, we present for the first time a recipe for one to quickly estimate corrections to the inspiral, ringdown, and remnant BH properties given only an arbitrary spacetime metric $g_{\alpha\beta}^\X$.
We exemplify this for both the JP and mod.-$\Delta$ spacetime metrics $g_{\alpha\beta}^\JP$ and $g_{\alpha\beta}^\MD$, which are parameterized by the single parameters $\epsilon_3$ and $\beta$, deviating from the Kerr metric $g_{\alpha\beta}^\K$.
We follow this up with a demonstration of the power of these corrections by performing the IMR consistency test to predict the magnitudes of $\epsilon_3$ and $\beta$ required for one to observe statistically significant deviations from the Kerr result.

Let us now briefly convey our primary findings.
Table~\ref{tab:results} presents a summary of the main results found in the following analysis.
Here we compare constraints on the JP and mod.-$\Delta$ deviation parameters $\epsilon_3$ and $\beta$ for each GW event and detector considered in this paper.
In particular, constraints are obtained using two different methods: (i) using the inspiral merger-ringdown consistency tests of GR in which one compares the inspiral and merger-ringdown signal's predictive power of the remnant BH mass and spin; and (ii) using the parameterized tests of GR, in which the Fisher analysis parameter estimation method is used to estimate the statistical uncertainties on template waveform parameters.
We find each method to agree very well with each other, and see that future detectors have the ability to constrain both $\epsilon_3$ and $\beta$ very stringently.
With current-generation GW detectors, we find comparable constraints on the JP deviation parameter $\epsilon_3$ to those from x-ray observations of BH accretion disks~\cite{Kong:2014wha,Bambi:2015ldr}, found to be loosely $\epsilon_3\lessapprox5$.
With future space-based and ground-based GW observatories, we find constraints a few orders of magnitude stronger.
We find that such results from the IMR consistency tests are mostly comparable to those from the parameterized tests. 
In particular, we find that the extreme-mass-ratio-inspirals observable by future space-based detector LISA~\cite{LISA} can probe such effects by three orders-of-magnitude stronger than the current constraints found in the literature.

Here we present the outline of this paper.
We present in Sec.~\ref{sec:waveform} how to find various corrections to the gravitational waveform in a generic way.
We follow this up in Sec.~\ref{sec:bumpy} with a review of the JP and mod.-$\Delta$ spacetime metrics considered in this analysis, along with corrections to the gravitational waveform.
In Sec.~\ref{sec:techniques} we lay out the gravitational waveform used in our analysis, as well as the Fisher analysis technique of parameter estimation, followed by an overview of the IMR consistency tests of GR and the parameterized tests.
Section~\ref{sec:results} presents our main results from these tests, and we finally offer concluding remarks and a discussion in Sec.~\ref{sec:conclusion}. 
Throughout this paper, we have adopted geometric units such that $G=1=c$.
In addition, we utilize the convention that $\dot{F}\equiv \frac{dF}{dt}$ and $F'\equiv\frac{dF}{dr}$, where additional dots and primes indicate additional consecutive derivatives.


\section{Corrections to gravitational waveforms}\label{sec:waveform}

In this section, we describe how modified BH solutions affect the gravitational waveform. In particular, we consider corrections to the inspiral, ringdown (through QNMs) and the final mass and spin of the remnant BH.

\subsection{Inspiral}

Among the many corrections to the gravitational waveform described in Sec.~\ref{sec:bumpy}, we consider the parameterized post-Einsteinian (ppE) formalism~\cite{Yunes:2009ke} for corrections to the inspiral phase and amplitude.
In this framework, the inspiral waveform in frequency domain $\tilde h$ can be described as
\begin{equation}\label{eq:ppE}
\tilde{h}_{\text{ppE}}=\mathcal{A}_{\GR}(f)(1+\alpha_\ppE u^{a_\ppE})e^{i[\Psi_{\GR}(f)+\beta_\ppE \; u^{b_\ppE}]},
\end{equation}
where $\Psi_{\GR}$ is the GR phase and $u=(\pi \mathcal{M} f)^{1/3}$ is the effective relative velocity of the compact objects with GW frequency $f$. $\mathcal{M}\equiv M_t \eta^{3/5}$ is the chirp mass with the total mass $M_t = m_1+m_2$  and the symmetric mass ratio $\eta\equiv m_1m_2/M_t^2$, and $m_A$ being the mass of the $A$th body.
The ppE parameters $\alpha_\ppE$ $(\beta_\ppE)$ determine the magnitude of the amplitude (phase) modifications to the waveform entering at $a_\ppE$ $(b_\ppE)$ powers of $u$~\footnote{The parameters $a_\ppE,b_\ppE$ can be related to the \emph{post-Newtonian} (PN) order $n$ by $a_\ppE=2n$ and $b_\ppE=2n-5$. Terms entering the waveform at $n$PN order are proportional to $(u/c)^{2n}$ relative to the leading-order term.}.

We now describe how to compute the ppE parameters for a given metric, following and slightly modifying App.~A of~\cite{Tahura_GdotMap}.
The calculation below is similar to that in~\cite{Cardenas-Avendano:2019zxd}, but has been extended for a more generic correction in the metric.
In particular, for the two example metrics that we consider in this paper, the dominant modifications to the binary evolution comes from the correction to the $(t,t)$ component of the metric, i.e. the Newtonian potential. First, we make an assumption that such a metric component is given by
\begin{equation}
g_{tt} = -1 + \frac{2M}{r}\left( 1 + A \frac{M^p}{r^p} \right) + \mathcal{O}\left(\frac{M^2}{r^2}\right),
\end{equation}
where $M$ is the mass of a BH and the parameters (A,p) characterize the leading correction to the potential. Then, the reduced effective potential of a binary becomes
\begin{equation}
\label{eq:V_eff}
V_\mathrm{eff} = - \frac{M_t}{r}\left( 1 + A \frac{M_t^p}{r^p} \right) + \frac{L_z^2}{2 \mu^2 r^2},
\end{equation}
where $\mu$ is the reduced mass while $L_z$ is the $z$-component of the orbital angular momentum. Taking the radial derivative of $V_\mathrm{eff}$ with respect to $r$, equating it with 0 and setting $L_z = \mu r^2 \Omega$ with the orbital angular velocity $\Omega$, one finds the modified Kepler's law as
\begin{equation}
\Omega^2 = \frac{M_t}{r^3}\left[ 1 +  (p+1) A \frac{M_t^p}{r^p} \right].
\end{equation}
This equation can be inverted to yield
\begin{equation}
\label{eq:r}
r = \left( \frac{M_t}{\Omega^2} \right)^{1/3}\left(1 +  \frac{p+1}{3} A \; v^{2p} \right),
\end{equation}
where  $v = (M_t \Omega)^{1/3}$ is the relative velocity and we only keep to leading correction in $A$. We substitute this back into Eq.~\eqref{eq:V_eff} and find the binding energy as
\begin{equation}
E_b = - \frac{1}{2} \mathcal{M} u^2 \left[1 - \frac{2(2p-1)}{3} A \; v^{2p} \right].
\end{equation}Next, we look at corrections to the GW luminosity. To take into account such dissipative corrections, one needs a specific theory. Thus, we neglect such effects in this paper and assume that the the GW luminosity is given by the one in GR~\cite{Cardenas-Avendano:2019zxd}:
\begin{equation}
\mathcal{L}_\GW = \frac{32}{5} \pi^6 \mu^2 r^4 f^6. 
\end{equation}
This luminosity acquires a conservative correction from that in Kepler's law as
\begin{equation}
\mathcal{L}_\GW = \frac{32}{5}  \eta^2 v^{10} \left[1 + \frac{4(p+1)}{3} A \; v^{2p} \right].
\end{equation}

Having these ingredients at hand, we are now ready to compute the ppE parameters. We first look at the frequency evolution of the binary, given by
\begin{eqnarray}
\label{eq:fdot}
\dot f = \frac{df}{dE_b} \frac{dE_b}{dt} &=& - \frac{df}{dE_b} \mathcal{L}_\GW \nonumber \\
&=&\frac{96}{5 \pi \mathcal{M}^2} u^{11} \left( 1 + \gamma_{\dot f} u^{2p}\right),
\end{eqnarray}
where $f = \Omega/\pi$ and 
\begin{equation}
\gamma_{\dot f} = \frac{2}{3}  \frac{(p+1) (2 p+1)}{\eta^{2p/5}} A.
\end{equation}
Equation~(20) of~\cite{Tahura_GdotMap} gives one the ppE parameters in the phase as\footnote{When we substitute $A=-4 a_1$ and $p=2$ where $a_1$ is the non-Kerr parameter used in~\cite{Rezzolla:2014mua,Cardenas-Avendano:2019zxd}, one finds $\beta_\ppE$ in agreement with that in~\cite{Cardenas-Avendano:2019zxd} modulo a minus sign that originates from the different convention used for the phase $\Psi$.}
\begin{eqnarray}
\beta_\ppE &=& - \frac{15}{16(2p-8)(2p-5)} \gamma_{\dot f} \nonumber \\
&=& -\frac{5  (p+1) (2 p+1)}{8 (2 p-8) (2 p-5)} \frac{A}{\eta^{2p/5}}, \\
b_\ppE &=& 2p -5. 
\end{eqnarray}
On the other hand, the amplitude correction can be obtained from Eqs.~\eqref{eq:r} and~\eqref{eq:fdot} and the fact that the amplitude is proportional to $r^2/\sqrt{\dot f}$:
\begin{eqnarray}
\alpha_\ppE &=&  -\frac{1}{3}  (p+1) (2 p-1) \frac{A}{\eta^{2p/5}}, \\
a_\ppE &=&  2p.
\end{eqnarray}
Notice that $\alpha_\ppE$ and $\beta_\ppE$ are related to each other as
\begin{equation}
\alpha_\ppE = \frac{16  (p-4) (2 p-5) (2 p-1)}{15 (2 p+1)} \beta_\ppE.
\end{equation}
Both corrections enter at $p$th PN order relative to the leading contribution in GR (or Kerr).
These expressions are generic and can be applied to any beyond-Kerr metrics, as long as the dominant correction to the metric comes from the correction to the Newtonian potential.

\subsection{Ringdown}

We next explain how to derive modifications to the ringdown portion of the waveform.
Following in the footsteps of the \emph{post-Kerr} formalism developed in Refs.~\cite{Silva:2019scu,Glampedakis:2017dvb,Glampedakis:2019dqh}, we estimate the QNM ringdown and damping frequencies $\omega_\R$ and $\omega_\I$ in the eikonal limit.
In this limit, $\omega_\R$ and $\omega_\I$ are associated with the light ring's angular frequency $\Omega_0$ and the Lyapunov exponent $\gamma_0$ (corresponding to the divergence rate of photon orbits grazing the light ring) at the light ring's radius $r_0$ as
\begin{eqnarray}
\omega_\R &=& 2 \Omega_0 =  2 (\Omega_\K + \delta \Omega_0),  \\
\omega_\I &=&  - \frac{1}{2}|\gamma_0| =  - \frac{1}{2}|\gamma_\K + \delta \gamma_0|.
\end{eqnarray}
Here 
\begin{equation}
\Omega_\K = \pm \frac{M^{1/2}}{r_\K^{3/2} \pm a M^{1/2}}
\end{equation}
is the angular frequency of the Kerr light ring\footnote{The upper (lower) sign corresponds to prograde (retrograde) orbit.} at 
\begin{equation}
r_\K = 2M \left\{ 1+\cos\left[ \frac{2}{3} \cos^{-1} \left(\mp \frac{a}{M} \right)\right] \right\},
\end{equation}
while 
\begin{align} 
\delta \Omega_{0} =&\mp\left(\frac{M}{r_{\K}}\right)^{1 / 2}\left[h_{\varphi \varphi} \pm\left(\frac{r_{\K}}{M}\right)^{1 / 2}\left(r_{\K}+3 M\right) h_{t \varphi}\right. \nonumber \\ 
&\left.+\left(3 r_{\K}^{2}+a^{2}\right) h_{t t}\right] /\left[\left(r_{\K}-M\right)\left(3 r_{\K}^{2}+a^{2}\right)\right] 
\end{align}
is the correction to $\Omega_\K$ with $h_{\mu\nu}$ representing the metric deviation away from Kerr. On the other hand,
\begin{equation}
\gamma_\K = 2 \sqrt{3 M} \frac{\Delta_{\K} \Omega_{\K}}{r_{\K}^{3 / 2}\left(r_{\K}-M\right)}
\end{equation}
is the Lyapunov exponent for Kerr with $\Delta_\K = r_\K^2 - 2 M r_\K + a^2$, while $\delta \gamma_0$ is the non-Kerr correction given in Eq.~(18) of~\cite{Glampedakis:2017dvb}.
See Refs.~\cite{McManus:2019ulj,Cardoso:2019mqo} where a general formalism to map ringdown corrections similar to the ones explained above directly to specific theories of gravity was developed.

\subsection{Final Mass and Spin}

Finally, we discuss modifications to the remnant BH's mass and spin, $M_f$ and $\chi_f$.
In GR, one can approximately estimate such parameters from the initial masses and spins via the specific energy $\tilde E$ and specific orbital angular momentum $\tilde{L}_\text{orb}$~\cite{Johannsen:2015pca}
\begin{align}
\label{eq:Eorb}\tilde{E}&=-\frac{g_{tt}+g_{t\phi}\Omega}{\sqrt{-g_{tt}-2g_{t\phi}\Omega-g_{\phi\phi}\Omega^2}}\\
\label{eq:Lorb}\tilde{L}_\text{orb}&=\pm\frac{g_{t\phi}+g_{\phi\phi}\Omega}{\sqrt{-g_{tt}-2g_{t\phi}\Omega-g_{\phi\phi}\Omega^2}}
\end{align}
of a particle of mass $\mu=m_1m_2/M_t$ orbiting the remnant BH at the inner-most-stable-circular-orbit (ISCO).
This corresponds to solving the equations~\cite{Buonanno:2007sv,Barausse:2009uz} 
\begin{align}
\nonumber \mu [1-\tilde E(M_f,\chi_f,r_\ISCO)]&=M_t-M_f,\\
\mu \tilde{L}_\text{orb}(M_f,\chi_f,r_\ISCO)&=M_t(M_f \chi_f-a_s-\delta_m a_a),
\end{align}
where $\delta_m\equiv(m_1-m_2)/M_t$ is the weighted mass difference, $a_{s,a}\equiv \frac{1}{2}(m_1\chi_1 \pm m_2\chi_2)$, and $r_\ISCO$ is the location of ISCO of the final BH. 
Since $\tilde E(M_f,\chi_f,r_\ISCO)$ is dimensionless in the geometric units, the $M_f$ dependence cancels and it only depends on $\chi_f$. 
Given the difference between $M_t$ and $M_f$ is small, we approximate $M_f \approx M_t$ in the second equation~\cite{Buonanno:2007sv}.
We estimate corrections to $M_f$ and $\chi_f$ in the beyond-Kerr metrics assuming this picture still holds.
The specific orbital energy and angular momentum are obtained such that the expressions $\bar V_\text{eff}=0$ and $\bar V_\text{eff}'=0$ are simultaneously satisfied for effective potential $\bar V_\mathrm{eff}$ given as~\cite{Johannsen:2015pca}
\begin{equation}\label{eq:Veff}
\bar{V}_\text{eff}=-\frac{\mu^2}{2}\left( g^{tt}E^2-2g^{t\phi} \tilde{E} \tilde{L}_\text{orb}+g^{\phi\phi} \tilde{L}_\text{orb}^2+1\right).
\end{equation}
Corrections to the ISCO radius are further obtained by solving the expression $\tilde{E}'(r_\ISCO)=0$ where $\tilde{E}$ is given by Eq.~\eqref{eq:Eorb}~\cite{Johannsen:2015pca}.
Combining these, one can find expressions for the corrections to the remnant black hole's mass $(\delta M_f)$ and spin $(\delta\chi_f)$ as
\begin{equation}
M_f =M_f^\K+\delta M_f, \quad
\chi_f=\chi_f^\K+ \delta\chi_f, 
\end{equation}
where $M_f^\K$ and $\chi_f^\K$  are the results for Kerr, which we take to be the ones in~\cite{PhenomDII}. See App.~\ref{appendix} for the general expressions for $\delta M_f$ and $\delta\chi_f$ in an arbitrary spacetime with metric $g_{\alpha\beta}^\X=g_{\alpha\beta}^\K+\zeta h_{\alpha\beta}^\X$.


\section{Beyond Kerr spacetimes}\label{sec:bumpy}
In this section we discuss the two beyond Kerr spacetimes considered in this analysis: the Johannsen-Psaltis metric introduced in Ref.~\cite{Johannsen:2011dh} and a modified version of Johannsen's metric in Refs.~\cite{Johannsen:2015mdd,Johannsen:2015pca}, denoted as the modified-$\Delta$ metric. Both of these are based on the Kerr metric whose components in Boyer-Lindquist coordinates are given by
\begin{align}
\nonumber g_{tt}^\K&=-\left( 1-\frac{2Mr}{\Sigma} \right) = - \frac{\Delta-a^2 \sin^2\theta}{\Sigma},\\ 
\nonumber g_{rr}^\K&=\frac{\Sigma}{\Delta},\\
\nonumber g_{\theta\theta}^\K&=\Sigma,\\
\nonumber g_{\phi\phi}^\K&=\left( r^2+a^2+\frac{2Ma^2r\sin^2\theta}{\Sigma} \right)\sin^2\theta\nonumber \\
&= \frac{ [(r^2+a^2)^2-a^2\Delta \sin^2\theta]\sin^2\theta}{\Sigma}, \nonumber \\
g_{t\phi}^\K&=-\frac{2Mar\sin^2\theta}{\Sigma} = - \frac{a (r^2+a^2-\Delta)\sin^2\theta}{\Sigma}, 
\label{eq:Kerr_metric}
\end{align}
with
\begin{align}
\nonumber \Sigma&\equiv r^2+a^2\cos^2\theta,\\
\Delta&\equiv r^2-2Mr+a^2,
\end{align}
where $(r,\theta)$ are the radial and polar coordinates, and $M$, $a$ are the BH's mass and spin.

We begin with an introduction to each spacetime, followed by the theoretical framework developed in the current analysis used to calculate the various non-Kerr corrections to the binary system present in each spacetime following Sec.~\ref{sec:waveform}.

\subsection{Johannsen-Psaltis metric}\label{sec:JP}
We begin our discussion on the JP metric, introduced by Johannsen and Psaltis in Ref.~\cite{Johannsen:2011dh}.
In this article, the authors begin with the Kerr metric $g_{\mu\nu}^\K$ in Eq.~\eqref{eq:Kerr_metric} and introduced a generalized parametric deviation $h(r,\theta)$ of the form
\begin{equation}
h(r,\theta)=\sum\limits_{k=0}^{\infty}\left(\epsilon_{2k}+\epsilon_{2k+1}\frac{Mr}{\Sigma}\right) \left( \frac{M^2}{\Sigma} \right)^k
\end{equation}
for some non-Kerr deviation parameters $\epsilon_k$ into each metric element.
By further applying the constraint of asymptotically flat spacetime at radial infinity, as well as observational constraints on the parameterized post-Newtonian framework~\cite{Will:2005va}, the deviation function $h(r,\theta)$ was reduced to a single non-Kerr parameter $\epsilon_3$
\begin{equation}
h(r,\theta)=\epsilon_3\frac{M^3r}{\Sigma^2}.
\end{equation}
Assuming that deviations from Kerr are small and keeping only up to linear order in $\epsilon_3$, the resulting JP metric $g_{\mu\nu}^\JP$ can be written as~\cite{Glampedakis:2017dvb}
\begin{align}
\nonumber g_{tt}^\JP&=-\left(1-\frac{2Mr}{\Sigma} \right)-\epsilon_3\frac{M^3 (r-2 M)}{r^4},\\ 
\nonumber g_{rr}^\JP&=\frac{\Sigma}{\Delta}+\epsilon_3\frac{M^3(r-2M)}{\Delta ^2},\\
\nonumber g_{\theta\theta}^\JP&=\Sigma,\\
\nonumber g_{\phi\phi}^\JP&=\left( r^2+a^2+\frac{2Ma^2r\sin^2\theta}{\Sigma} \right)\sin^2\theta+\epsilon_3\frac{a^2M^3(r+2M)}{r^4},\\
 g_{t\phi}^\JP&=-\frac{2Mar\sin^2\theta}{\Sigma}-\epsilon_3\frac{2aM^4}{r^4}.\label{eq:JP}
\end{align}
With this choice of $h(r,\theta)$, the JP metric now allows one to probe strong-field gravity to any order of spin in a parameterized way.
Observe how in the limit of $\epsilon_3\rightarrow0$, we recover the original Kerr metric for a spinning BH.
See Refs.~\cite{Kong:2014wha,Bambi:2015ldr} for constraints on the JP deviation parameter $\epsilon_3$ from BH accretion disk thermal spectra, found to be loosely $\epsilon_3\lessapprox5$.

We next identify the dominant contribution to the binary evolution. 
For a particle orbiting around a BH, the angular velocity $\Omega$ is determined from the radial derivative of $g_{tt}$, $g_{t\phi}$ and $g_{\phi\phi}$~\cite{Johannsen:2015pca} as
\begin{equation}\label{eq:kepler}
\Omega=\frac{-\partial_r\pm\sqrt{(\partial_r g_{t\phi})^2-\partial_rg_{tt}\partial_rg_{\phi\phi}}}{\partial_r g_{\phi\phi}}.
\end{equation}
When we expand the JP metric components about $r =\infty$, one finds that the leading correction to $\partial_r g_{tt}$, $\partial_r g_{t\phi}$ and $\partial_r g_{\phi\phi}$ enters at $\mathcal{O}(M^2/r^2)$, $\mathcal{O}(M^3/r^3)$ and $\mathcal{O}(M^5/r^5)$ relative to the leading Kerr contribution respectively. Thus, the dominant correction comes from $g_{tt}$ and we find 
\begin{equation}
A_\JP = - \frac{\epsilon_3}{2}, \quad p_\JP = 2. 
\end{equation}

Now let us lay the groundwork for the JP modifications to the gravitational waveform by applying the results presented in Sec.~\ref{sec:waveform}. 
First, the ppE parameters are given by
\begin{align}
\nonumber \beta_\ppE^\JP&=\frac{75 \epsilon_3}{64 \eta ^{4/5}}, \hspace{5mm} b_\ppE^\JP=-1,\\ \label{eq:JPppe}
\alpha\ppE^\JP&=\frac{3 \epsilon_3}{2 \eta ^{4/5}}, \hspace{5mm} a_\ppE^\JP=4,
\end{align}
and the corrections enter at 2PN order.
This is of the same order as the correction for the beyond-Kerr metric proposed in Ref.~\cite{Rezzolla:2014mua}, as found in Ref.~\cite{Cardenas-Avendano:2019zxd}.
Next, the QNM corrections in a JP spacetime to first order in JP deviation parameter, and quadratic in spin are given by
\begin{align}
\nonumber \omega_\R^\JP &= \omega_\R^\K+ \epsilon_3\left( \frac{1}{81\sqrt{3}M}+\frac{10}{729M}\chi+\frac{47}{1458\sqrt{3}M}\chi^2 \right),\\
\omega_\I^\JP &= \omega_\I^\K- \epsilon_3\left( \frac{1}{486M}\chi+\frac{16}{2187\sqrt{3}M}\chi^2 \right),
\end{align}
for unitless spin parameter $\chi\equiv a / M$, and Kerr QNM frequencies $\omega_{\R,\I}^\K$. 
It is interesting to note that $\omega_\I$ does not acquire corrections if the BH is non-spinning.
Finally, corrections to the final mass and spin are given by
\begin{widetext}
\begin{align}
\delta M_f^\JP&=-\epsilon_3\frac{ \mu}{139968} \left\lbrack864 \delta\chi_f^\JP \left(5 \sqrt{2} \chi_f^\K+3 \sqrt{3}\right)+545 \sqrt{2} \left( \chi_f^\K \right)^2+324 \sqrt{3} \chi_f^\K+216 \sqrt{2}\right\rbrack,\\
\nonumber\delta \chi_f^\JP &= -\epsilon_3 \frac{1}{384 \sqrt{3} \kappa  \mu} \Big\lbrack420 \mu  M_t \chi _a \delta _m+420 \mu  \chi _a \lambda-152 \sqrt{2} \kappa  \mu +2416 \sqrt{3} \mu ^2+420 \mu  \delta _m \lambda \chi _s\\
&\hspace{4.5cm}+945 \sqrt{3} M_t^2-315 \kappa  M_t+1086 \sqrt{6} \mu  M_t+420 \mu  M_t \chi _s\Big\rbrack,
\end{align}\\
which is valid to linear order in $\epsilon_3$ and to quadratic order in the final spin, and
\begin{eqnarray}
\kappa&\equiv& \sqrt{8 \sqrt{3} \mu  \chi _a \left(M_t \delta _m+\lambda\right)+8 \sqrt{3} \mu  \chi _s \left(\delta _m \lambda+M_t\right)+3 \left(40 \mu ^2+9 M_t^2+12 \sqrt{2} \mu  M_t\right)}, \\
\lambda&\equiv&\sqrt{M_t(M_t-4\mu)}.
\end{eqnarray}
\end{widetext} 
This is derived from $r_\ISCO$, which, to linear order in JP deviation and quadratic in BH spin, is given by
\begin{equation}
r_\ISCO^\JP=r_\ISCO^\K\left\lbrack 1-\epsilon_3\left(\frac{1}{27} + \frac{37}{324\sqrt{6}}\chi +\frac{1229}{23328}\chi^2 \right) \right\rbrack,
\end{equation}
with Kerr result $r_\ISCO^\K$~\cite{wald_1984}.

\subsection{modified-$\Delta$ metric}\label{sec:MD}

Now let us discuss the newly constructed mod.-$\Delta$ metric, following in the footsteps of Johannsen in Refs.~\cite{Johannsen:2015mdd,Johannsen:2015pca}.
We begin in Ref.~\cite{Johannsen:2015pca}, in which 4 free functions $A_1(r)$, $A_2(r)$, $A_5(r)$, and $f(r)$ are introduced to the Kerr spacetime,  parameterically describing deviations from GR, as shown in Eq.~(51) of~\cite{Johannsen:2015pca}.
Such a metric is found to be stationary, axisymmetric, asymptotically flat, admits freely-rotating BHs, reduces to the Kerr metric for $A_1(r)=A_2(r)=A_5(r)=1$, and $f(r)=0$, and possess a third constant of motion, a Carter-like constant~\cite{Carter:1968rr}.
This symmetry, as in the Kerr metric, gives rise to separable, non-chaotic geodesic equations for particle motion.

Following this, in Ref.~\cite{Johannsen:2015mdd}, Johannsen further modified the obtained spacetime metric  by introducing a pure-deviation $\beta$ from the Kerr metric, by substituting
\begin{equation}
\label{eq:Delta_replacement}
\Delta\rightarrow \bar{\Delta}\equiv\Delta+\beta M^2
\end{equation}
into the metric found in Eq.~(51) of Ref.~\cite{Johannsen:2015pca}.
We further equate all other free functions to their Kerr values, $A_1(r)=A_2(r)=A_5(r)=1$, and $f(r)=0$, resulting in the modified-$\Delta$ metric $g_{\mu\nu}^\MD$ with elements given by\footnote{This metric can also be obtained by applying the replacement in Eq.~\eqref{eq:Delta_replacement} to Eq.~\eqref{eq:Kerr_metric}.}
\begin{align}
\nonumber g_{tt}^\MD&=-\left( 1-\frac{2Mr}{\Sigma} \right)-\beta \frac{M^2}{\Sigma},\\ 
\nonumber g_{rr}^\MD&=\frac{\Sigma}{\Delta}-\beta\frac{M^2\Sigma}{\Delta^2},\\
\nonumber g_{\theta\theta}^\MD&=\Sigma,\\
\nonumber g_{\phi\phi}^\MD&=\left( r^2+a^2+\frac{2Ma^2r\sin^2\theta}{\Sigma} \right)\sin^2\theta\\
\nonumber& \hspace{5mm} -\beta\frac{a^2M^2\sin ^4\theta}{\Sigma},\\
g_{t\phi}^\MD&=-\frac{2Mar\sin^2\theta}{\Sigma}+\beta\frac{a M^2  \sin ^2\theta}{\Sigma},\label{eq:MD}
\end{align}
where we assume that the deviation from Kerr is small and we keep only to linear order in $\beta$.
This spacetime is entirely parameterized by the single, pure-deviation parameter $\beta$, reduces to the Kerr metric for $\beta=0$, and is useful as it can be mapped to BH solutions other than Kerr.
Such metrics include the Kerr-Newman metric for charged BHs~\cite{delaCruzDombriz:2009et}, the RS-II braneworld BHs~\cite{Aliev:2005bi}, and those in the modified gravity (MOG)~\cite{Moffat:2014aja}.

Now let us consider the various corrections to the gravitational waveform present in the mod.-$\Delta$ spacetime. Just like in the case of the JP metric, the leading correction comes from $g_{tt}^\MD$ and 
\begin{equation}
A_\MD = - \frac{\beta}{2}, \quad p_\MD = 1,
\end{equation}
which means that the correction enters at 1PN order\footnote{The $\beta$-correction enters at $\mathcal{O}(M^2/r^2)$ order higher than the leading in $\Delta$, and indeed, the leading correction to $g_{tt}$ also enters at $\mathcal{O}(M^2/r^2)$ order higher than the leading contribution. However, such a leading term is a constant and the $\beta$-correction enters only at $\mathcal{O}(M/r)$ order higher than the leading Newtonian potential.}
First, the ppE parameters entering in the inspiral waveform are given by
\begin{align}
\nonumber \beta_\ppE^\MD&=\frac{5\beta}{48\eta^{2/5}}, \hspace{5mm} b_\ppE^\MD=-3,\\
\alpha_\ppE^\MD&=-\frac{\beta}{3\eta^{2/5}}, \hspace{5mm} a_\ppE^\MD=2.
\end{align}
Next, the ringdown frequencies are modified as
\begin{align}
\nonumber \omega_\R^\MD &= \omega_\R^\K+ \beta \left( \frac{1}{9\sqrt{3}M}+\frac{2}{27M}\chi+\frac{61}{486\sqrt{3}M}\chi^2 \right),\\
\omega_\I^\MD &= \omega_\I^\K+\beta \left( \frac{1}{108\sqrt{3}M}-\frac{1}{243M}\chi-\frac{11}{729\sqrt{3}M}\chi^2 \right).
\end{align}
Finally, the corrections to the final mass and spin are given by 
\begin{widetext}
\begin{align}
\delta M_f^\MD&=-\beta\frac{\mu}{7776}\left\lbrack48 \delta\chi_f^\MD \left(5 \sqrt{2} \chi_f^\K+3 \sqrt{3}\right)+119 \sqrt{2} \left(\chi_f^\K\right)^2+84 \sqrt{3}\chi_f^\K+72 \sqrt{2}\right\rbrack, \\
\nonumber\delta\chi_f^\MD&=-\beta\frac{1}{256 \sqrt{3} \kappa  \mu}\Big\lbrack876 \mu  M_t \chi _a \delta _m+876 \mu  \chi _a \lambda -286 \sqrt{2} \kappa  \mu +5288 \sqrt{3} \mu ^2+876 \mu  \delta _m \lambda \chi _s\\
&\hspace{4.5cm}+1971 \sqrt{3} M_t^2-657 \kappa  M_t+2172 \sqrt{6} \mu  M_t+876 \mu  M_t \chi _s\Big\rbrack,
\end{align}
\end{widetext}
which are valid to linear order in $\beta$ and quadratic order in $\chi_f^\K$.
We used the ISCO radius expression of
\begin{equation}
r_\ISCO^\MD=r_\ISCO^\K \left\lbrack 1-\beta \left( \frac{1}{4} + \frac{1}{2\sqrt{6}}\chi + \frac{77}{432}\chi^2 \right) \right\rbrack,
\end{equation}
which is valid to quadratic order in spin.


\section{Parameter-estimation techniques}\label{sec:techniques}

In this section we introduce the technical methods and formalisms utilized in this analysis.
In particular, we discuss the gravitational waveform, the Fisher analysis parameter estimation method, the IMR consistency tests of GR, the parameterized tests of GR and finally the detectors and GW events considered. 

\subsection{Gravitational waveform}
Let us begin by discussing our gravitational waveform template.
We utilize the non-precessing, sky-averaged \emph{IMRPhenomD} GR waveform obtained via the NR fits of Refs.~\cite{PhenomDI,PhenomDII}.
Typically, the IMRPhenomD waveform is parameterized in terms of the $(\mathcal{M},\eta,\chi_a,\chi_s)$ mass and spin parameters, where $\chi_{s,a} = (\chi_1 \pm \chi_2)/2$ is the symmetric and anti-symmetric dimensionless spins.
However, in this analysis we instead re-parameterize it by computing the expressions for $\mathcal{M}(M_f,\eta,\chi_s,\chi_f,\zeta)$ and  $\chi_a(M_f,\eta,\chi_s,\chi_f,\zeta)$, where $M_f$ and $\chi_f$ are the remnant BH's mass and spin that include corrections from Kerr computed in the following section. $\zeta = \epsilon_3 \ \mathrm{or} \ \beta$ represents the deviation parameter from Kerr.
By substituting in the above expressions, we obtain the IMRPhenomD waveform parameterized instead by the $(M_f,\eta,\chi_s,\chi_f)$ mass and spin parameters, allowing us to directly generate multi-dimensional posterior probability distributions between the final mass and spin $M_f$ and $\chi_f$.
The resulting template waveform consists of
\begin{equation}
\theta^a=\left( \ln{\mathcal{A}_\GR},\phi_c,t_c,M_f,\eta,\chi_f,\chi_s,\zeta \right),
\end{equation}
where $\mathcal{A}_\GR\equiv\frac{\mathcal{M}_z^{5/6}}{\sqrt{30}\pi^{2/3}D_L}$ is a generalized, sky-averaged amplitude in GR with redshifted chirp mass $\mathcal{M}_z\equiv\mathcal{M}(1+z)$ for redshift $z$, $D_L$ is the luminosity distance, while $\phi_c$ and $t_c$ are the coalescence phase and time.

We then modify this IMRPhenomD waveform by including corrections explained in Secs.~\ref{sec:waveform} and~\ref{sec:bumpy}. Namely, we first modify the inspiral portion by introducing the ppE parameters. We next modify the ringdown and damping frequencies. For the Kerr contribution to these frequencies, we use those given in~\cite{PhenomDII}. Finally, we endow corrections to the final mass and spin.

\subsection{Parameter estimation}\label{sec:PE}
Now let us discuss the parameter estimation method utilized in this investigation.
Similar to Refs.~\cite{Carson:2019kkh,Carson_QNMPRD,Carson_QNM_PRL} by the same authors, we use a Fisher-based analysis to estimate the final mass and spin posterior probability distributions used in the IMR consistency test.
Such an analysis is not as robust as the comprehensive Bayesian one used in e.g.~\cite{Ghosh_IMRcon,Ghosh_IMRcon2,Abbott_IMRcon,Abbott_IMRcon2}, though it is useful for order-of-magnitude parameter estimations, without the significant time constraint of a Bayesian analysis.
This topic was thoroughly investigated in Ref.~\cite{Yunes_ModifiedPhysics} and additionally Refs.~\cite{Carson_multiBandPRD,Zack:Proceedings} by the same authors, where it was found that for loud enough events, the results well approximate their Bayesian counterpart.
At both $2$PN (for the JP metric) and $1$PN (for the MD metric) orders, the former reference found very strong agreement between the two methods for a GW150914-like event with a signal-to-noise-ratio (SNR) of $25.1$, with such agreements only strengthening considerably for the future detectors with loud events considered in this analysis. As we state in Sec.~\ref{sec:conclusion}, there are several caveats in our analysis. Accordingly, the goal of this paper is to find an order-of-magnitude estimate on current and future bounds on beyond-Kerr parameters. For this aim, Fisher results are sufficient.

\renewcommand{\arraystretch}{1.2}
\begin{table*}
\centering
\resizebox{\linewidth}{!}{%
\addvbuffer[12pt 8pt]{\begin{tabular}{c | c c c c c c c c}
\multirow{2}{*}{Event} & \multirow{2}{*}{$m_1$ $\lbrack M_\odot\rbrack$} & \multirow{2}{*}{$m_2$ $\lbrack M_\odot\rbrack$} & \multirow{2}{*}{$\chi_1$} & \multirow{2}{*}{$\chi_2$} & \multirow{2}{*}{$D_L$ $\lbrack$Mpc$\rbrack$} & SNR & $f_\text{low}$ $\lbrack$Hz$\rbrack$ & $f_\text{high}$ $\lbrack$Hz$\rbrack$  \\
&&&&&& (O2, CE, CE+LISA) & (O2, CE, LISA) & (O2, CE, LISA)\\
\hline 
GW150914~\cite{GW150914} & 35.8 & 29.1 & $0.32$ & $-0.44$ & 410 & $(25.1,\text{ }1930,\text{ }1940)$ & $(23,1,0.02)$ & $(5\times10^3,\text{ }5\times10^3,\text{ }1)$\\
GW170729~\cite{GW170729} & 51.0 & 31.9 & $0.60$ & $-0.57$ & 2,850 & $(10.7,\text{ }405,\text{ }410)$ & $(23,1,0.01)$ & $(5\times10^3,\text{ }5\times10^3,\text{ }1)$ \\
EMRI & $10^6$ & $10$ & $0.90$ & $-0.50$ & 3,000 & 18.7 & $10^{-3}$ & $1$\\
SMBHB & $10^6$ & $5\times10^{4}$ &  $0.90$ & $-0.90$ & 3,000 & 1830 & $6\times10^{-5}$ & $1$\\
\end{tabular}}
}
\caption{List of GW events considered in the following analysis, along with their masses $m_{1,2}$, dimensionless spins $\chi_{1,2}$, the luminosity distance $D_L$ and SNR (for aLIGO O2, CE, CE+LISA), and finally the upper and lower cutoff frequencies $f_\text{low,high}$. For the latter quantity, we list the frequencies for the aLIGO O2, CE, and LISA detectors respectively. For the first two events, we list the luminosity distance corresponding to the central values measured by the LIGO/Virgo Collaborations, though we scaled them appropriately so that the SNRs become the ones listed here using the waveform and the noise curves described in this paper. For the final two events EMRI and SMBHB, the SNR and frequency cutoffs are listed only for LISA.
}\label{tab:events}
\end{table*}

Here we briefly introduce the Fisher analysis method of parameter estimation discussed thoroughly in Refs.~\cite{Cutler:Fisher,Poisson:Fisher,Berti:Fisher,Yagi:2009zm}.
The statistical uncertainties on template parameters $\theta^a$ can be shown to be approximately
\begin{equation}
\Delta\theta^i=\sqrt{\tilde{\Gamma}_{ii}^{-1}},
\end{equation}
with effective Fisher matrix $\tilde{\Gamma}_{ij}=\Gamma_{ij}+(\sigma_{\theta^i}^{(0)})^{-2}\delta_{ij}$.
$\sigma_{\theta^i}^{(0)}$ represent the root-mean-square uncertainties for the prior probability distributions (assumed to be Gaussian~\cite{Cutler:Fisher,Poisson:Fisher}), and the \emph{Fisher information matrix} is defined as
\begin{equation}
\Gamma_{ij}\equiv \left( \frac{\partial h}{\partial \theta^i} \Bigg| \frac{\partial h}{\partial \theta^j} \right).
\end{equation}
Here the inner product $(a|b)$ is weighted by the detector noise spectral density $S_n(f)$ like so
\begin{equation}
(a|b)\equiv2\int^{f_\text{high}}_{f_\text{low}}\frac{\tilde{a}^*\tilde{b}+\tilde{b}^*\tilde{a}}{S_n(f)}df,
\end{equation}
with $f_\text{high,low}$ representing the detector-dependent upper and lower cutoff frequencies, as tabulated for each event considered in this analysis in Table~\ref{tab:events}.
In particular, for ground-based detectors the lower cutoff frequencies are given by the detector-dependent values of 23 Hz and 1 Hz for aLIGO O2 and CE, while the upper cutoff frequencies are chosen such that the GW spectrum is sufficiently small compared to the detector sensitivity $S_n(f)$.
For space-based detectors, the lower frequency is chosen to be the frequency four years prior to merger $f_\text{4yrs}$ (corresponding to LISA's conservative mission lifetime) as can be found e.g. in Ref.~\cite{Berti:Fisher}.
Finally, if one further desires to combine the detections from multiple detectors with Fisher matrices $\bm{\Gamma}^\text{A}$ and $\bm{\Gamma}^\text{B}$, the effective Fisher matrix can be shown to be
\begin{equation}
\tilde{\Gamma}^\text{tot}_{ij}=\Gamma^\text{A}_{ij}+\Gamma^\text{B}_{ij}+\frac{1}{(\sigma_{\theta^i}^{(0)})^2}\delta_{ij}.
\end{equation}

Reference~\cite{Cutler:2007mi} shows how one can estimate the ``theoretical", or systematic errors present in the extraction of template parameters $\theta^a$ due to mismodeling present in the template waveform.
In particular, one can approximate the systematic errors present in $\theta^a$ by assuming use of GR template with Kerr BHs, while an alternative spacetime is in fact the correct theory described by nature.
The resulting expression for systematic errors is given by
\begin{equation}
    \Delta_\text{th} \theta^a\approx\left( \Gamma^{-1} \right)^{ab} \left( \left\lbrack \Delta A+i A_\GR\Delta\Psi \right\rbrack e^{i\Psi_\GR} | \partial_b h_\GR \right),
\end{equation}
where $A_\GR$, $\Psi_\GR$, and $h_\GR$ are the amplitude, phase, and waveform for Kerr BH binaries, and $\Delta A \equiv A_\GR-A_\zeta$ and $\Delta \Psi \equiv \Psi_\GR-\Psi_\zeta$ are the differences in amplitudes and phases between the Kerr and beyond-Kerr expressions.
In the following analysis, we steadily increase the value of $\zeta$ in the JP or mod.~$\Delta$ spacetimes to increase the systematic mismodeling uncertainties between it and the one in a Kerr spacetime.

Next we present the prior template probability distributions, and fiducial parameter values used in our analysis.
We impose Gaussian prior distributions with root-mean-square errors $\sigma_{\theta^a}^{(0)}$ given by $|\phi_c|\leq\pi$, $|\chi_s|\leq 1$, and $|\chi_f|\leq 1$.
We use fiducial template values such that $\eta$ and $\chi_s$ correspond to the initial parameters of the GW event being considered, $M_f$ and $\chi_f$ correspond to those predicted by the expressions computed in the following section, and $\phi_c=t_c=0$.

Finally, we present the specific detectors and GW events considered in this analysis.
Specifically, we regard current-generation ground-based detector aLIGO O2~\cite{aLIGO} (whose sensitivity is similar to that for the current run O3), the future ground-based detector Cosmic Explorer (CE)~\cite{Ap_Voyager_CE}, and finally the future space-based detector LISA~\cite{LISA}, with detector sensitivities all displayed in Ref.~\cite{Carson_multiBandPRD}.
In particular, we focus our attentions on the ``golden'' GW150914-like~\cite{GW150914} events, more massive GW170729-like~\cite{GW170729} events, EMRIs\footnote{Such EMRIs are not valid in the NR fits presented in the IMRPhenomD waveform, which have been calibrated to NR simulations with mass ratios of only up to 1:18. To take this into account in the parameterized tests, we remove all NR fits from the gravitational waveform, and cut-off all frequency integrations before the merger-ringdown, at $f_\ISCO$. Namely, we use the TaylorF2 waveform in GR up to 3.5PN order included in the phase and introduce the ppE corrections to account for the inspiral corrections. Thus, such estimates are more conservative than the other ones presented in this paper, as we only consider the inspiral portion of the waveform.}. and SMBHBs.
For the former two events, we also consider the multi-band detections combined with both CE and LISA to further enhance the number of GW cycles observed.
See Table~\ref{tab:events} for a comprehensive list of GW events considered in this analysis, in addition to their source properties (masses, spins, luminosity distances, SNRs, and detector cutoff frequencies).

\subsection{IMR consistency test}\label{sec:IMR}
Finally, we introduce the IMR consistency tests of GR~\cite{Ghosh_IMRcon,Ghosh_IMRcon2,Abbott_IMRcon,Abbott_IMRcon2,Zack:Proceedings} used to test our resolving power of non-Kerr metrics.
In accordance to the no-hair theorem of GR, the remnant BH formed from the coalescence of two BHs with masses and spins $m_\A$ and $\chi_\A$ can be described entirely by only its final mass $M_f(m_1,m_2,\chi_1,\chi_2)$ and final spin $\chi_f(m_1,m_2,\chi_1,\chi_2)$.
Each of these expressions can be computed using the NR fits found in Ref.~\cite{PhenomDII}.
If GR was indeed the true theory of gravity present in nature and BHs are Kerr, the final mass and spin parameters can be independently and accurately predicted using only the inspiral GW signal (I, $f<f_\ISCO=(6^{3/2}\pi M_t)^{-1}$), but also from the merger-ringdown signal (MR, $f>f_\ISCO$).
Alternatively, if another metric described the spacetime we occupy (say JP or mod.-$\Delta$), the $M_f$ and $\chi_f$ obtained from each portion of the GW signal would begin to disagree under the assumption that the compact objects are Kerr BHs.
In this investigation, we predict the size of JP or mod.-$\Delta$ deviations from GR necessary to become observable in the IMR consistency test as above.

Let us briefly present the application of the IMR consistency test, with a more thorough description left to Refs.~\cite{Carson_multiBandPRD,Zack:Proceedings,Carson_QNMPRD,Carson_QNM_PRL} by the same authors.
To begin, the two-dimensional Gaussian probability distributions $P_{\I,\MR}(M_f,\chi_f)$ between BH mass and spin parameters are estimated using the Fisher analysis techniques introduced in Sec.~\ref{sec:PE}, from each portion I and MR of the gravitational waveform individually.
For this step of the IMR consistency test, we choose $\zeta=0$ corresponding to a Kerr waveform.
Further, systematic uncertainty shifts in each remnant BH parameter $\bm{\Delta_\text{th}X_{\I,\MR}}\equiv(\delta_\text{th}M_f,\Delta_\text{th}\chi_f)$ are introduced as described in Sec.~\ref{sec:PE}.
Such systematic uncertainties are calculated via the difference (i.e. the waveform mismodeling) between the assumed Kerr waveform ($\zeta=0$), and the beyond-Kerr waveform ($\zeta\ne0$).
Finally, the I and MR probability distributions for the final-state variables $\bm{X}\equiv\{M_f,\chi_f\}$ can be written as
\begin{eqnarray}
P_{\I,\MR}(\bm X)&\equiv& \frac{1}{2\pi\sqrt{|\bm{\Sigma}_{\I,\MR}|}} \nonumber \\
&& \times \exp \left\lbrack -\frac{1}{2} \left(  \bm{X} - \bm{X}^\GR -\bm{\Delta_\text{th}X}_{\I,\MR} \right)^\text{T} \right. \nonumber \\
& & \left. \times \bm{\Sigma}_{\I,\MR}^{-1}\left( \bm{X} - \bm{X}^\GR -\bm{\Delta_\text{th}X}_{\I,\MR} \right) \right\rbrack,\label{eq:pdf}
\end{eqnarray}
with covariance matrices $\bm{\Sigma_{\I,\MR}}$ and the GR predictions $\bm{X}^\GR$ for the final state variables.
The resulting agreement with the GR values of $\bm{X}^\GR$, as computed by the NR fits of~\cite{PhenomDII} (simply speaking, comparing the statistical uncertainties to the systematic ones), as well as with the ones between I and MR, indicates the degree with which the acquired signal agrees with that predicted by GR.

Such probability distributions contain both statistical uncertainties $\sqrt{\bm{\Sigma}}$ deterministic of their \emph{size}, and systematic uncertainties $\bm{\Delta_\text{th}X_{\I,\MR}}$ indicating their offset from the GR values of remnant BH mass and spin.
In the following analysis, similar to that presented in Refs.~\cite{Carson_multiBandPRD,Zack:Proceedings,Carson_QNMPRD,Carson_QNM_PRL} by the same authors, we introduce each correction computed in the following section to the gravitational waveform, and then slowly vary the magnitudes of the JP and mod.-$\Delta$ deviation parameters $\zeta$.
The deviation parameters are increased until the $90\%$ posterior probability distributions obtained in each portion of the gravitational waveform begin to disagree with each other.
At this point, one could definitively approximate the magnitude of non-Kerr parameters required to present themselves as evidence of being a viable beyond-GR spacetime.

\subsection{Parameterized tests}\label{sec:param}
In this section we provide a brief overview of the parameterized tests of GR used to compare results from the IMR consistency test. 
The so-called parameterized test of GR is a useful tool that allows one to obtain upper-bounds on template parameters (e.g. $\zeta$) that parameterize the gravitational waveform beyond-Kerr in a simple way.
Typically, such tests rely only on inspiral corrections to the inspiral waveform via the ppE formalism~\cite{Yunes:2009ke}.
We extend this usual format by further including waveform corrections to the ringdown, and the remnant BH properties as described in Sec.~\ref{sec:waveform}.
All such corrections are entirely parameterized by the single non-Kerr parameter $\zeta$, providing for a simple test of the ``non-Kerr'' behavior of a given signal.
We then make use of the Fisher analysis techniques described in Sec.~\ref{sec:PE} with a fiducial value of $\zeta=0$ (Kerr).
The corresponding root-mean-square uncertainties on $\zeta$ effectively describe the amount of ``wiggle room'' the parameter can vary within, while still remaining consistent with the detector's noise.
We can therefore take such variation as an upper bound on the beyond-Kerr parameter $\zeta$.


\section{Results}\label{sec:results}

In this section we present the primary results obtained in this analysis, first in the JP spacetime, followed by the mod.-$\Delta$ spacetime.
See Table~\ref{tab:results} for a summary of all estimated constraints in both the JP and mod.-$\Delta$ spacetimes obtained from both (i) the IMR consistency tests of GR, and (ii) the parameterized tests of GR for comparison.

\begin{figure*}[!htbp]
\begin{center}
\includegraphics[width=.45\textwidth]{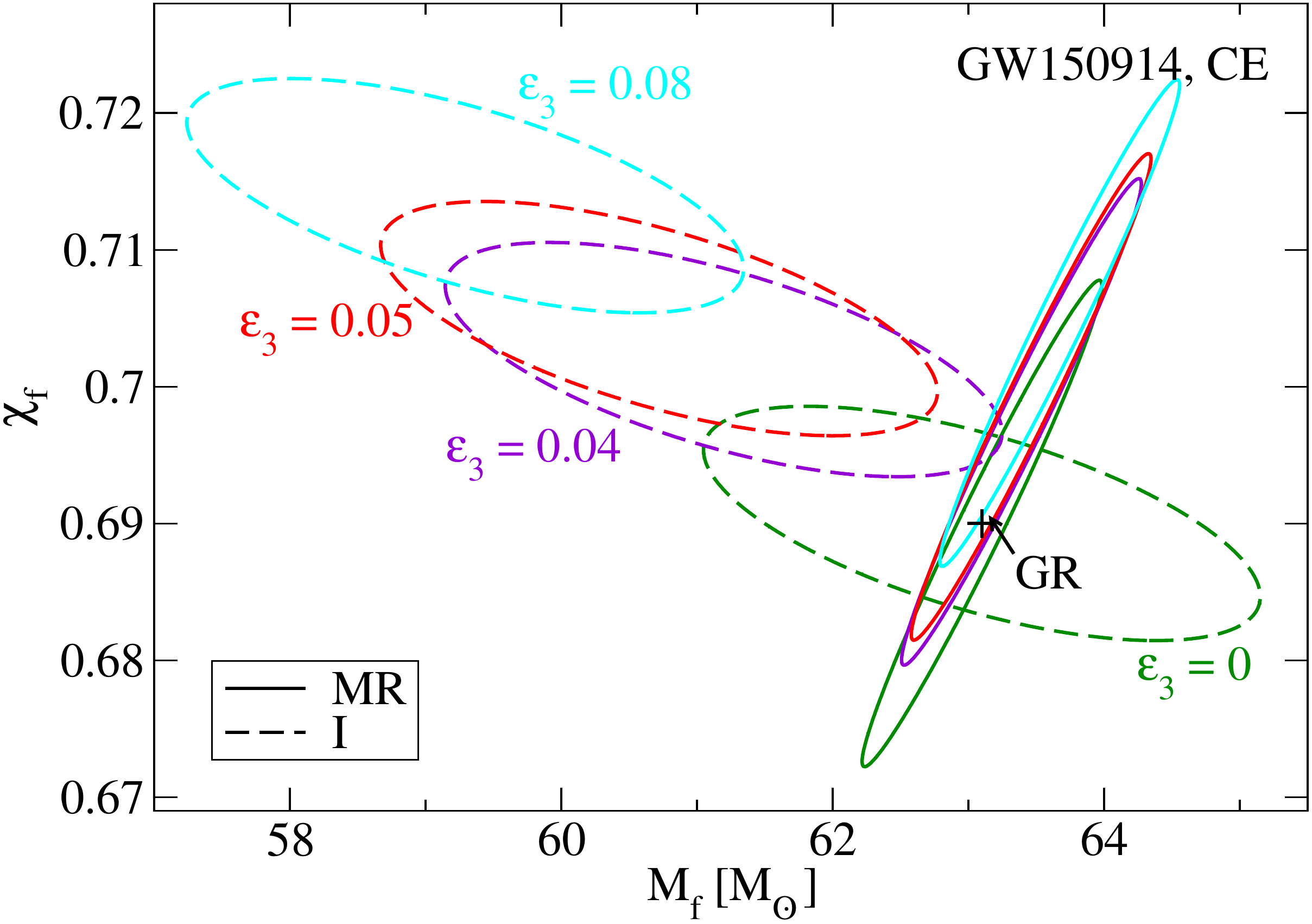}
\includegraphics[width=.45\textwidth]{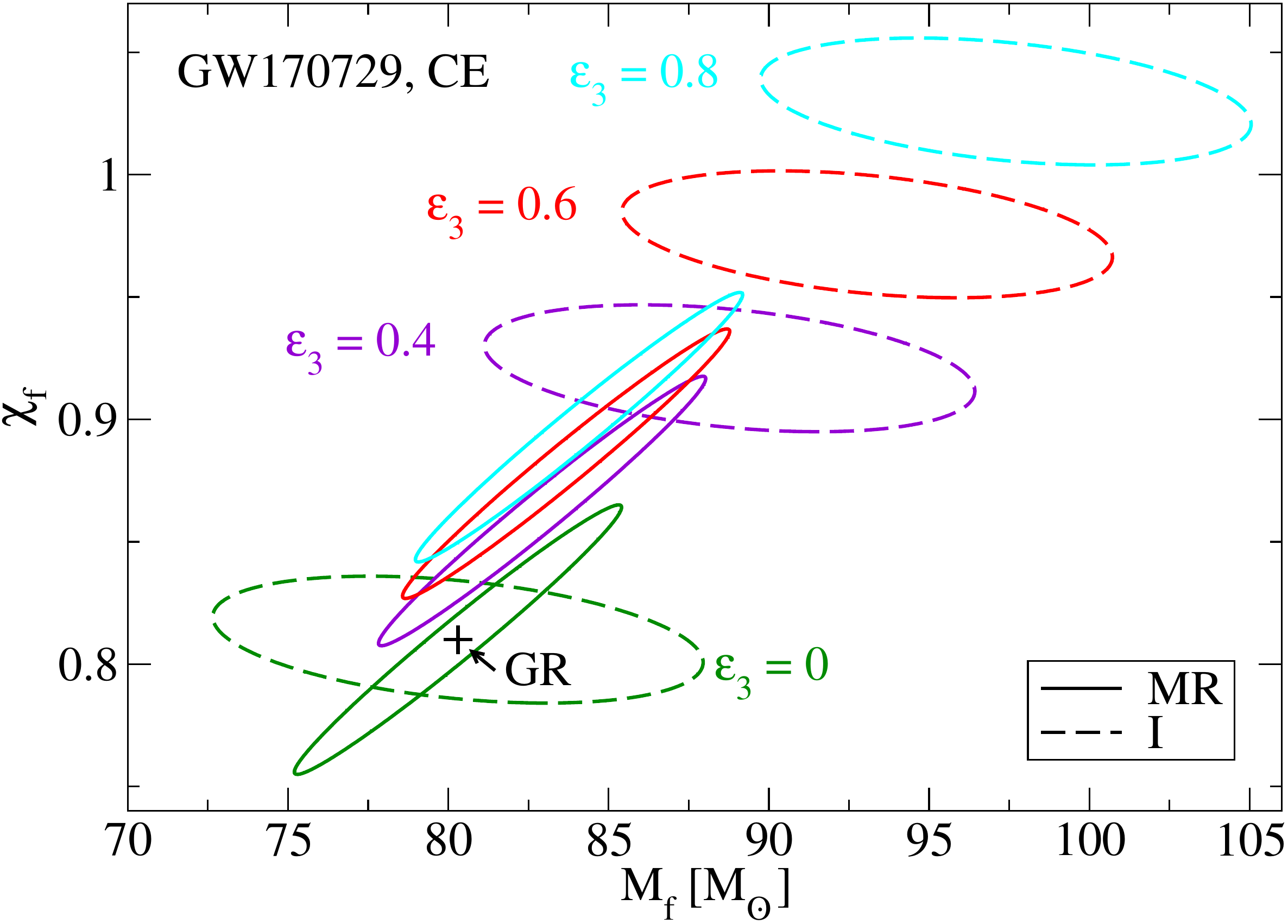}
\includegraphics[width=.45\textwidth]{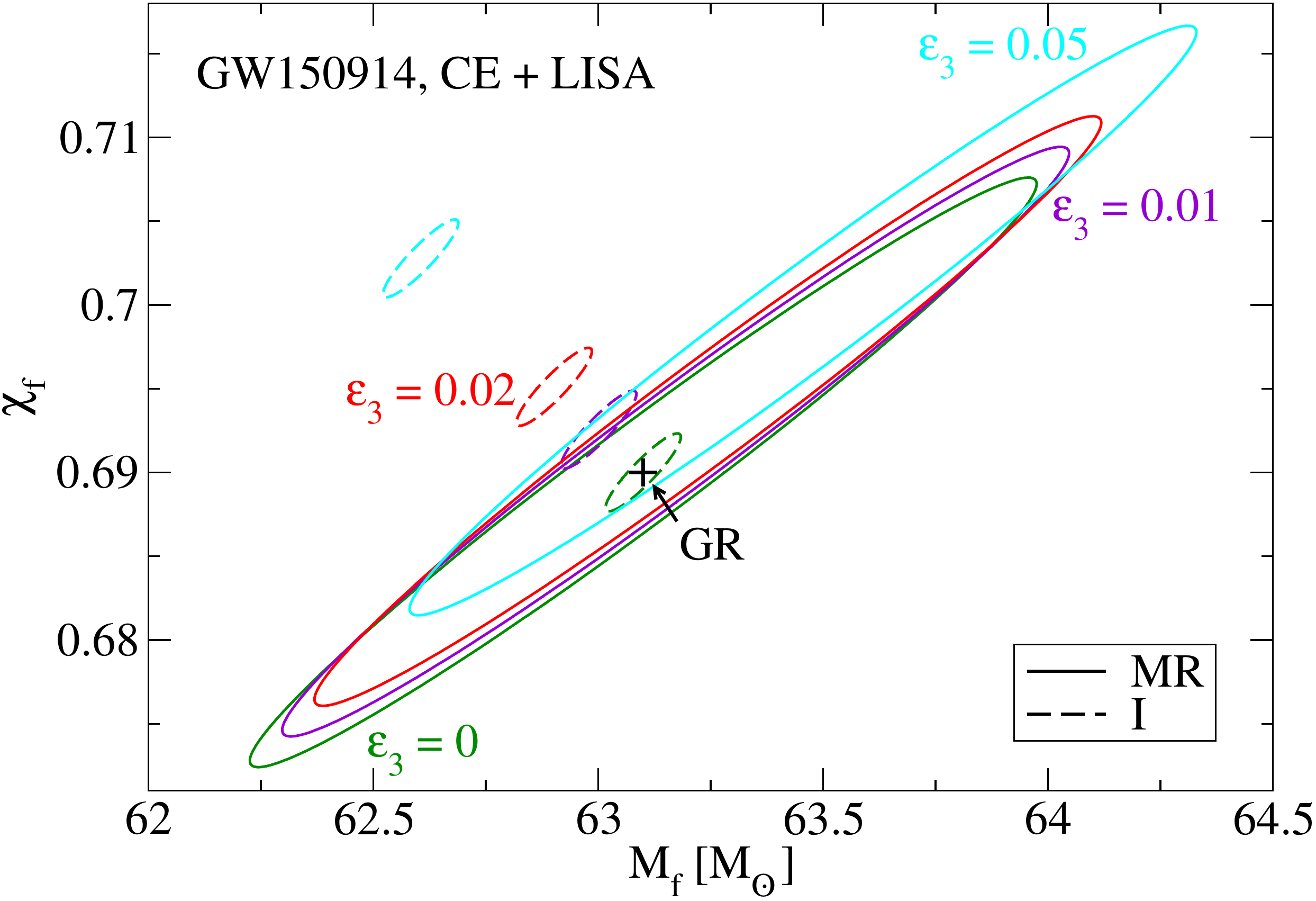}
\includegraphics[width=.45\textwidth]{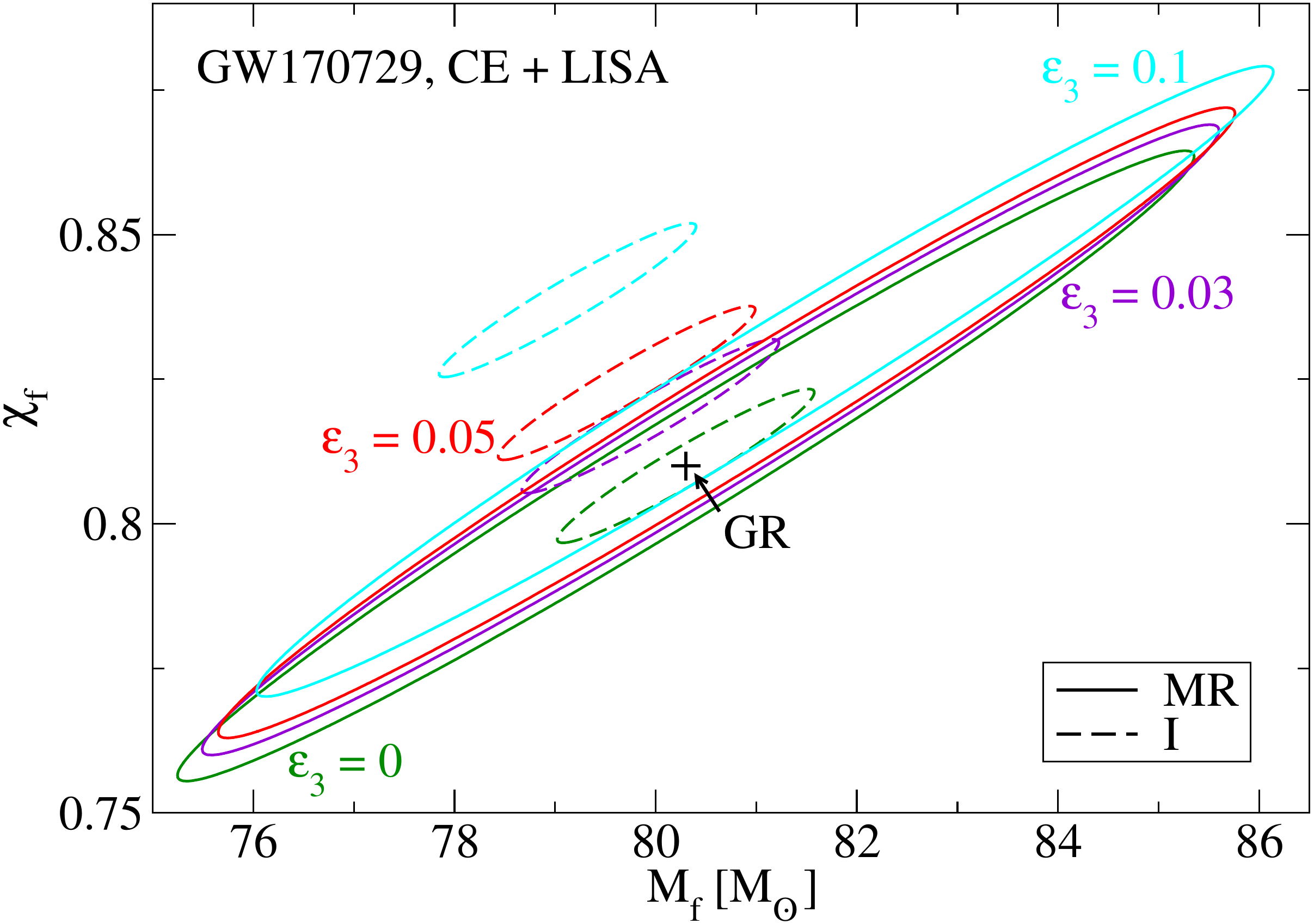}
\caption{IMR consistency test for the the ``golden event'' GW150914 (left) and the massive event GW170729 (right) in the JP spacetime using the CE detector (top), and through the multiband observation between CE and LISA (bottom).
In particular, in each panel we plot the 90\% confidence regions in the $(M_f,\chi_f)$ plane as observed from (i) only the inspiral (I) signal, and (ii) only the merger-ringdown (MR) signal, for consecutively increasing values of the JP deviation parameter $\epsilon_3$.
Only when such probability distributions begin to disagree with each other can one decisively admit there may be evidence of beyond-Kerr spacetimes present.
}\label{fig:imrJP}
\end{center}
\end{figure*}

\subsection{JP spacetime}\label{sec:IMRJP}

\subsubsection{IMR consistency tests}

Let us begin by performing the IMR consistency test in the JP spacetime, to predict how well one can observe deviations from GR.
By following the procedure outlined in Sec.~\ref{sec:IMR}, we perform the IMR consistency test for several consecutively increasing values of the JP deviation parameter $\epsilon_3$ until the inspiral and merger-ringdown 90\% confidence interval probability distributions begin to disagree.
Only then can one provide evidence of non-Kerr behaviors in the gravitational signal.

We start with an investigation into the GW events already detected on the aLIGO O2 detector, namely GW150914 and GW170729.
We perform the IMR consistency test for several values of $\epsilon_3$ injected into the gravitational waveform with the aLIGO O2 GW detector.
We find that for GW150914-like (GW170729-like) events, when $\epsilon_3\approx7$ ($\epsilon_3\approx10$) the systematic uncertainties begin to overtake the statistical errors, and the I and MR contours begin to disagree. 
Such constraints on $\epsilon_3$ are on the same order of magnitude as those from x-ray observations presented in~\cite{Kong:2014wha,Bambi:2015ldr}.
However, they fail to satisfy the small-deviation assumption made in the derivation of ppE parameters, thus the resulting constraints are less valid than the following ones presented for future GW detectors.
As a result of this we do not present the resulting contours in this paper, however the constraints are still tabulated in Table~\ref{tab:results} for reference.

We next focus our attention on future observations of the same GW events GW150914 and GW170729.
We now consider such events as detected by the future CE detector, as well as the increased observation from the multiband observation between ground- and space-based detectors CE and LISA.
Figure~\ref{fig:imrJP} displays the results of the IMR consistency test in such cases.
For the CE case, we find that when $\epsilon_3\approx0.05$ and $\epsilon_3\approx0.6$, we can begin to distinguish the inspiral and merger-ringdown signals for GW150914- and GW170729-like events respectively. 
Notice that the plot range is much smaller than that in the top row for aLIGO. This means that the error ellipses are much smaller for the CE case than the aLIGO case due to larger SNRs.
For the multiband case, we find that while the inspiral has significantly smaller statistical uncertainties than the merger-ringdown (due to the low-frequency space-based observations by LISA), its systematic uncertainties are much larger.
This allows one to constrain deviation parameters to $\epsilon_3\approx0.02$ and $\epsilon_3\approx0.05$ for GW150914- and GW170729-like events respectively.
Such constraints are about two-orders-of-magnitude stronger than the existing bounds presented by~\cite{Kong:2014wha,Bambi:2015ldr}.

\begin{figure}[!htbp]
\begin{center}
\includegraphics[width=.9\linewidth]{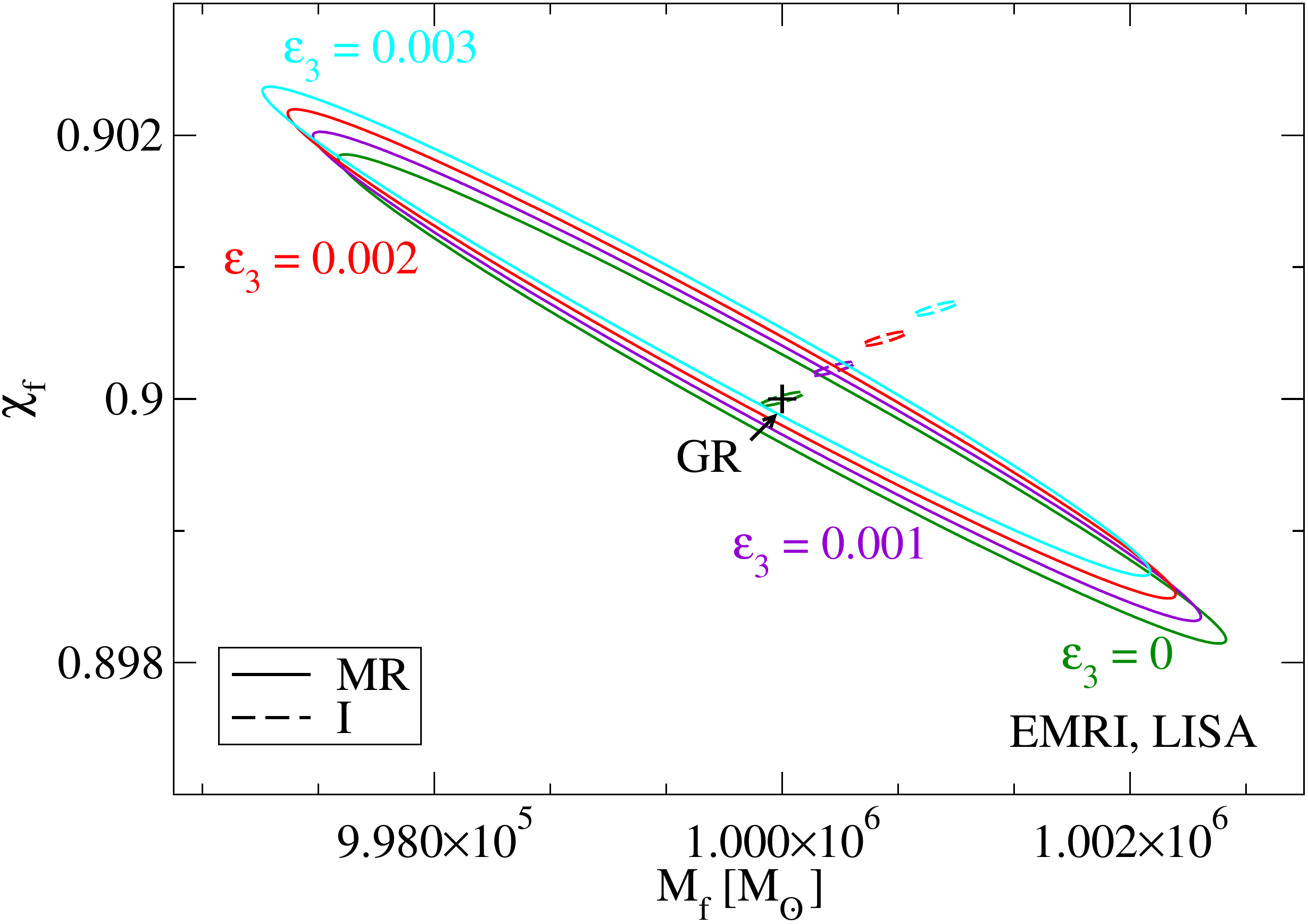}
\includegraphics[width=.9\linewidth]{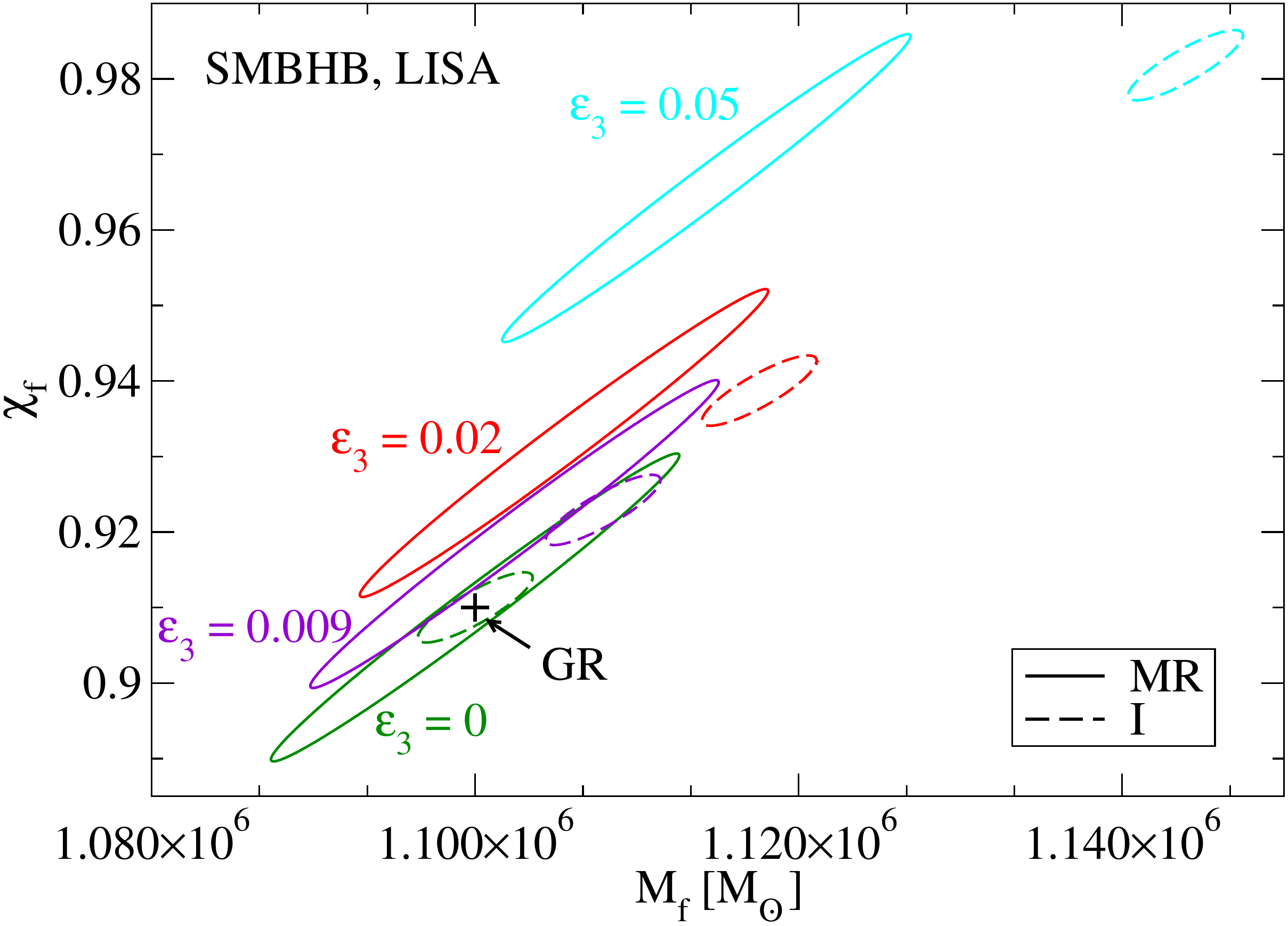}
\caption{Similar to Fig.~\ref{fig:imrJP}, but for the EMRI and SMBHB GW events.
}\label{fig:imrJPmassive}
\end{center}
\end{figure}

Finally, we consider the more extreme events detectable in the low frequency bands by LISA: EMRIs and SMBHBs.
Figure~\ref{fig:imrJPmassive} presents the resulting IMR consistency test for such two events involving massive BHs.
When considering EMRI systems, we find that the inspiral signal is very deterministic for the remnant BH properties, with such contours orders of magnitude smaller than their merger-ringdown counterparts. 
With a majority of systematic uncertainties present in the inspiral signal, we find that we can constrain $\epsilon_3\approx2\times10^{-3}$ -- a few orders-of-magnitude stronger than those found in~\cite{Kong:2014wha,Bambi:2015ldr}.
These constraints are much stronger because in high mass-ratio inspirals, the quadrupole radiation is smaller, thus the orbit decays slower and the number of GW cycles is greatly increased compared to equal-mass systems, so the non-Kerr effect is significantly enhanced. 
This can be seen by the factor of $\eta^{-4/5}$ present in the ppE phase parameter in Eq.~\eqref{eq:JPppe}, which is very large for large mass-ratio systems ($\sim 10^{4}$ for EMRIs, and only $\sim 3$ for i.e. GW150914).
However, as noted previously, we point out that such results are not as reliable due to the IMRPhenomD NR fits only being calibrated up to mass ratios of 1:18.
Finally, we see that for SMBHB events detected by LISA, we can constrain $\epsilon_3\approx0.02$, much weaker than those from EMRIs, and similar to those found by future GW170729 and GW150914 observations.

\subsubsection{Parameterized tests}

For comparison we perform a parameterized test of GR for the deviation parameter $\epsilon_3$.
To do so, we include $\epsilon_3$ into the template waveform with fiducial value of $0$, and perform a Fisher analysis to estimate root-mean-square uncertainties on $\epsilon_3$.
Such results are displayed in Table~\ref{tab:results} in comparison to all of the constraints found via the IMR consistency test as presented here.
We find that they give comparable bounds on $\epsilon_3$ for each case considered, even for EMRIs.
In this case, the IMRD consistency test is less valid as mentioned above, while in the parameterized test we used the TaylorF2 waveform with the ppE correction and stopped all integrations before the merger-ringdown.

\begin{figure*}
\begin{center}
\includegraphics[width=.45\textwidth]{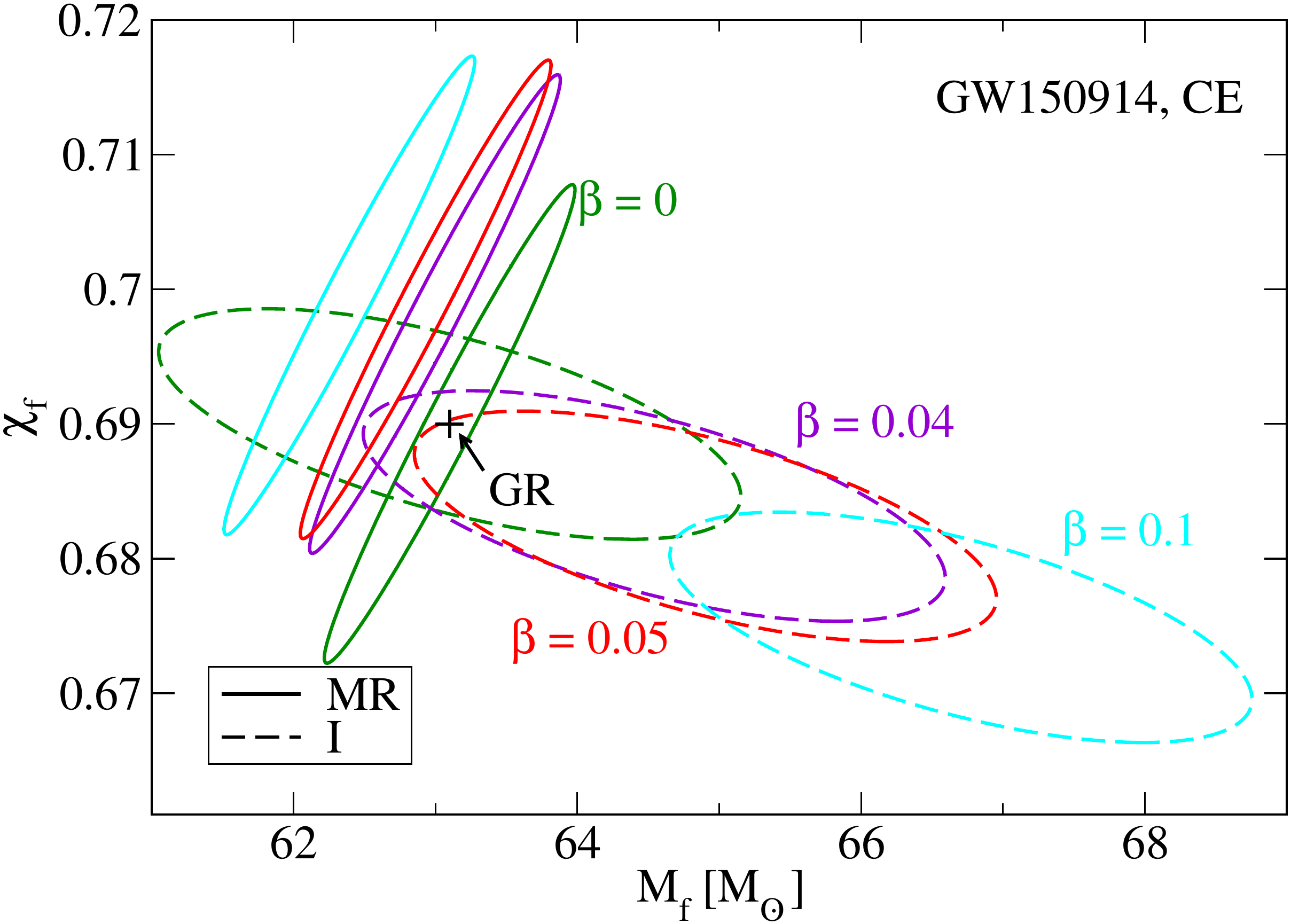}
\includegraphics[width=.45\textwidth]{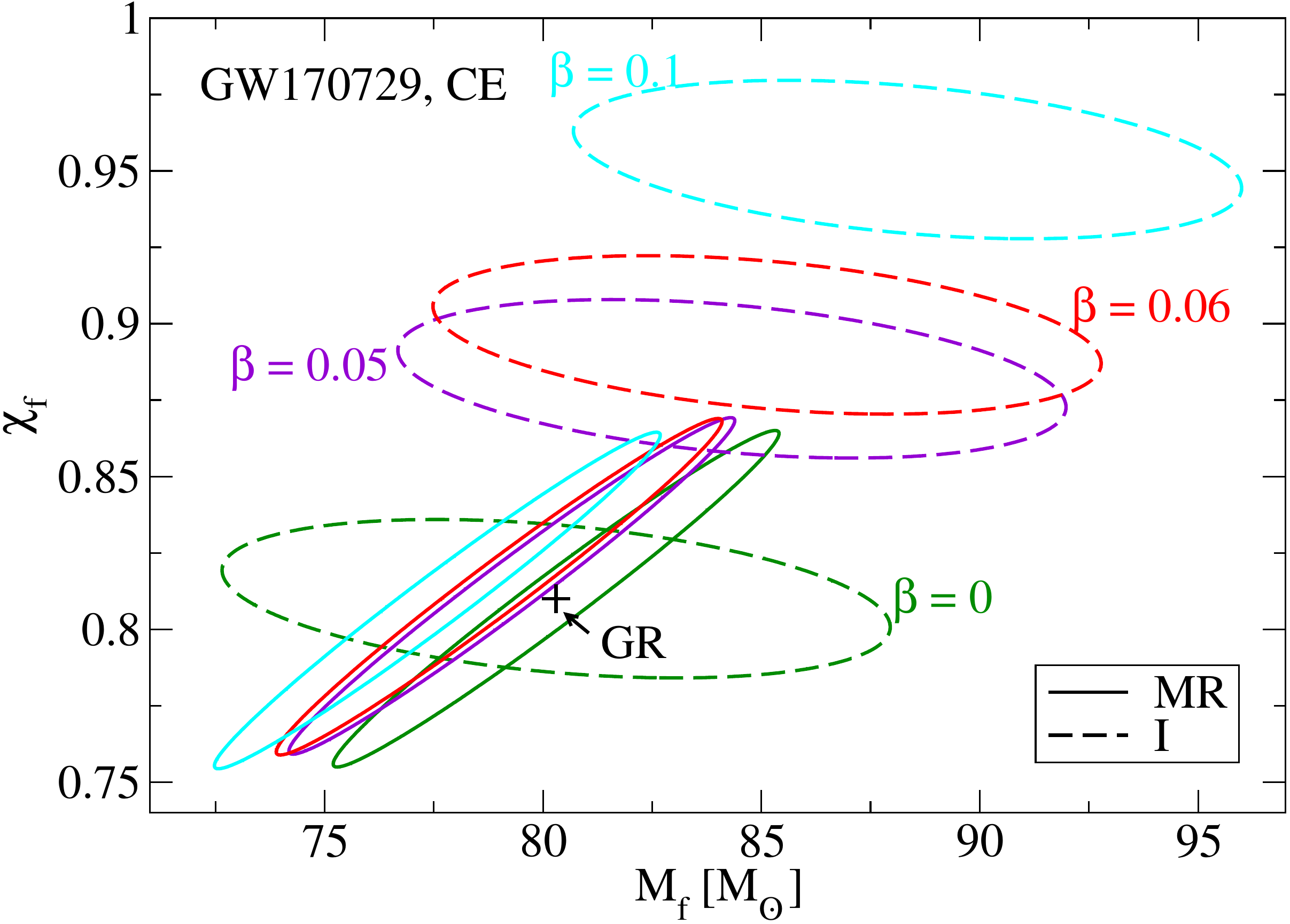}
\includegraphics[width=.45\textwidth]{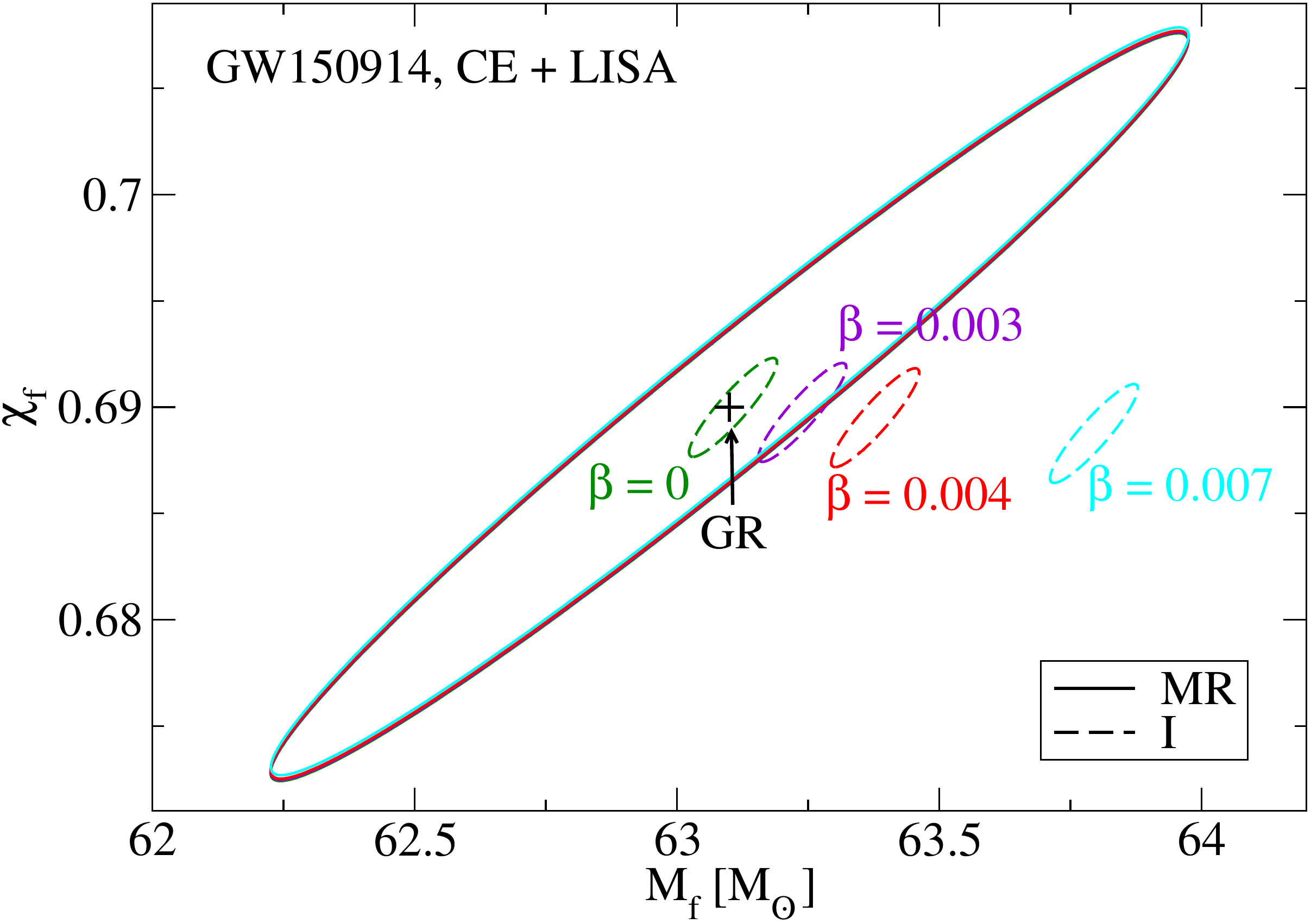}
\includegraphics[width=.45\textwidth]{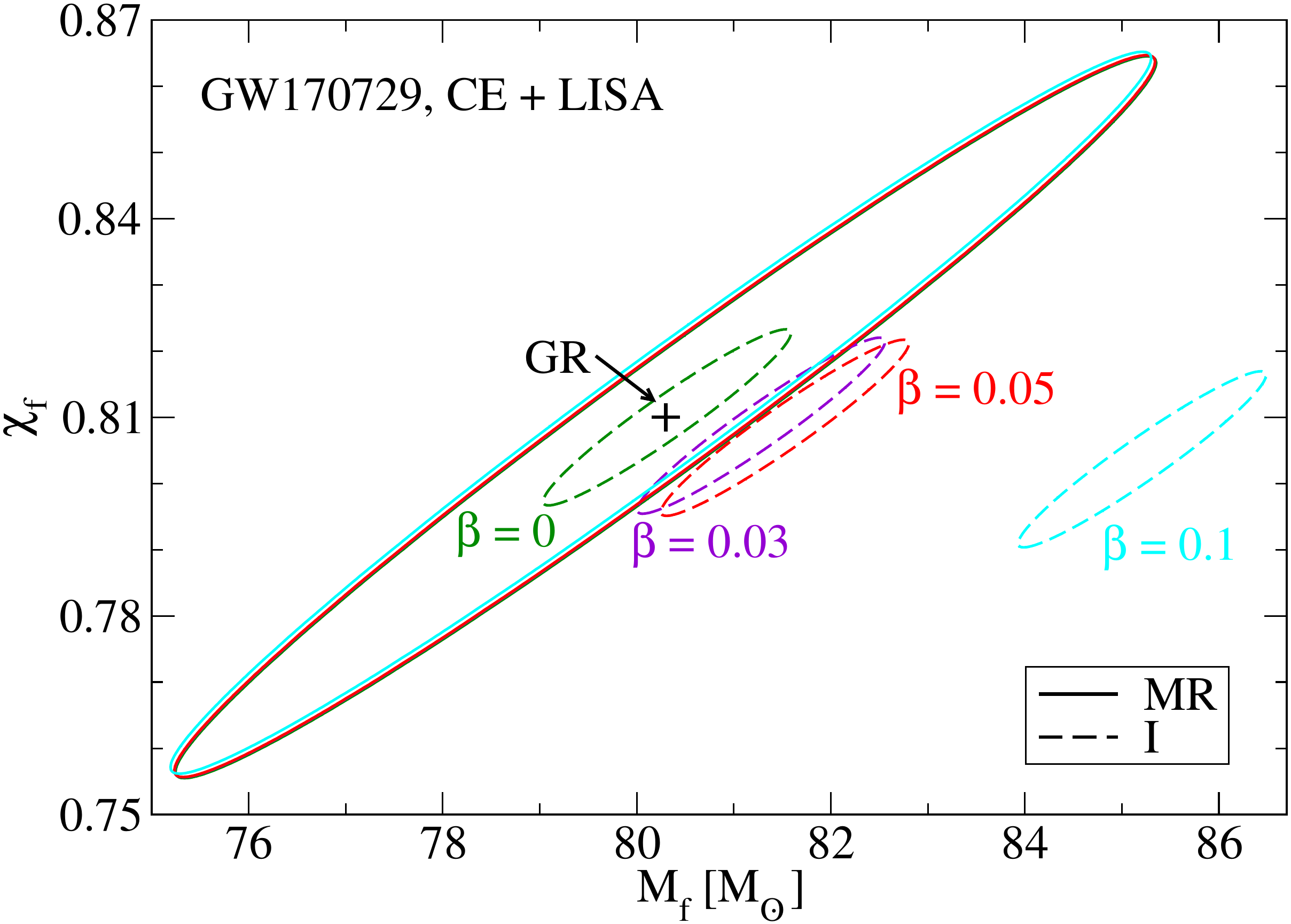}
\caption{
Same as Fig.~\ref{fig:imrJP} but under the mod.-$\Delta$ spacetime instead, with $\beta$ being the beyond-Kerr deviation parameter.
}\label{fig:imrKB}
\end{center}
\end{figure*}

\begin{figure}
\begin{center}
\includegraphics[width=.9\linewidth]{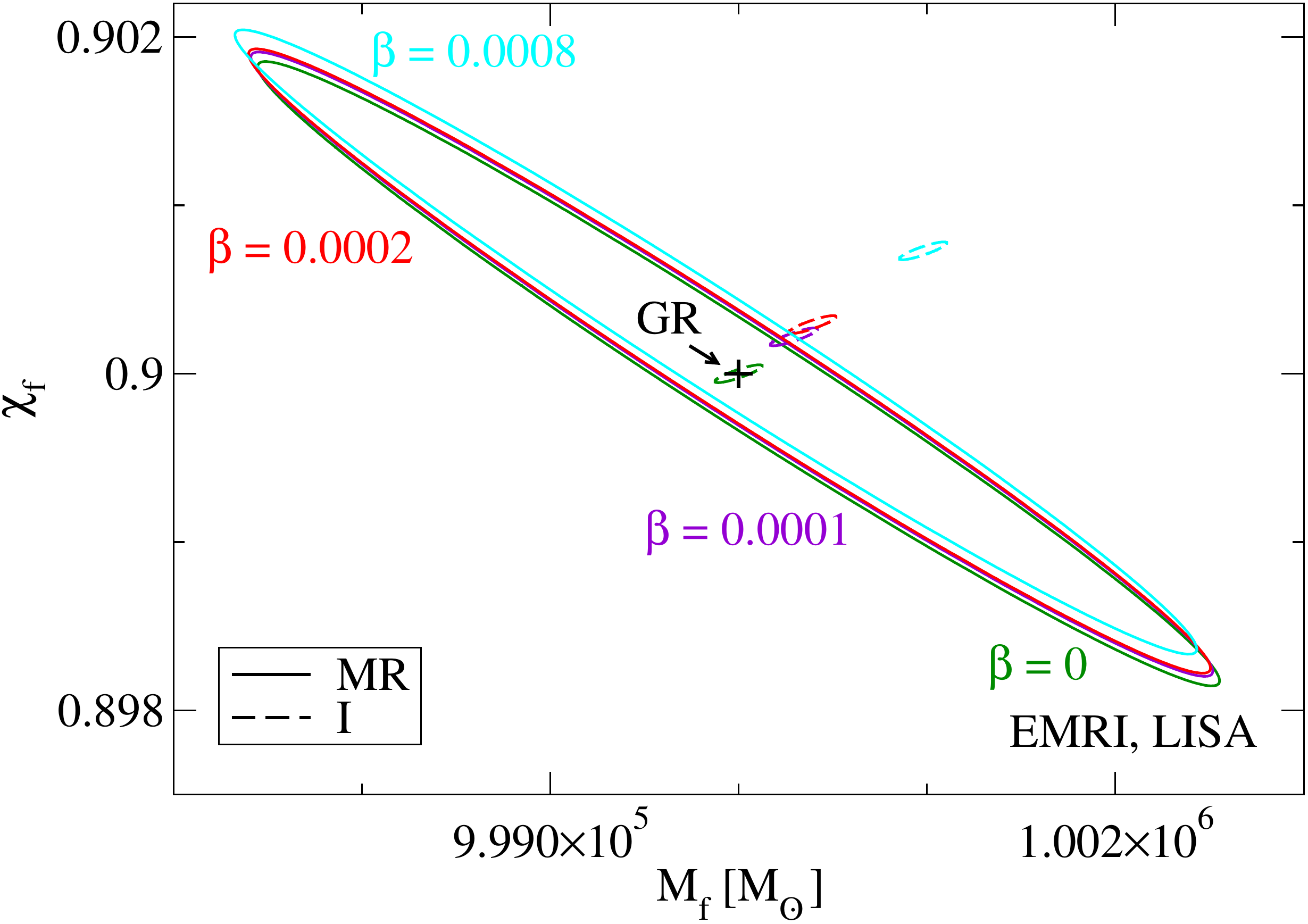}
\includegraphics[width=.9\linewidth]{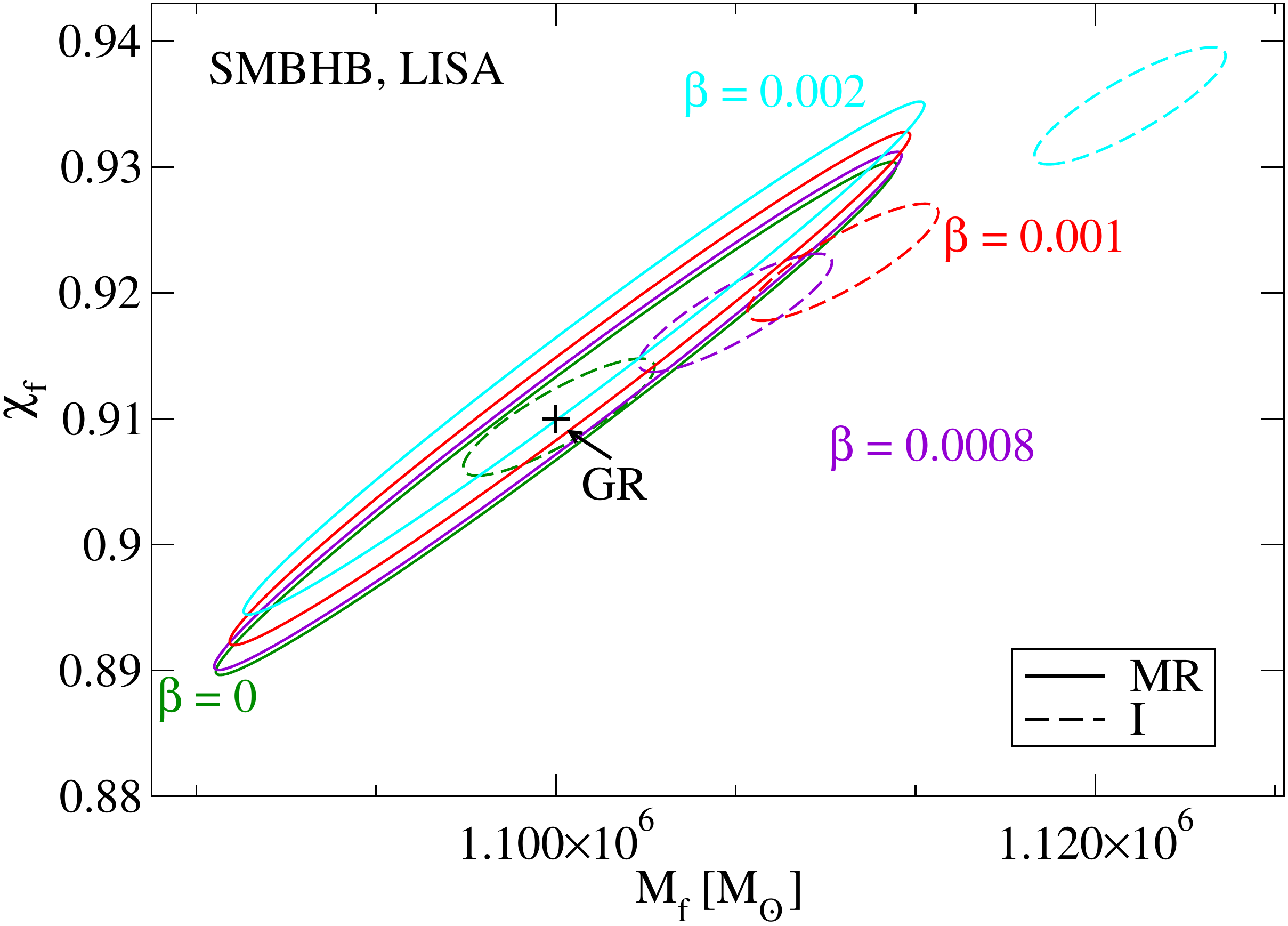}
\caption{
Same as Fig.~\ref{fig:imrJPmassive} but within the mod.-$\Delta$ spacetime with deviation parameter $\beta$.
}\label{fig:imrKBmassive}
\end{center}
\end{figure}

\renewcommand{\arraystretch}{1.2}
\begin{table}
        \centering
        \begin{tabular}{c | c c}
        & Inspiral-corr. only & all corrections \\ 
        \hline
        GW150914 (CE) & $0.0586$ & $0.0509$ \\
        GW170729 (CE) & $0.464$ & $0.406$ \\
        SMBH (LISA) & $0.0101$ & $0.0098$ \\
        \end{tabular}
        \caption{
        Comparison between constraints on JP parameter $\epsilon_3$ when (left) only inspiral corrections to the waveform are included, and (right) when all of the inspiral, ringdown, and remnant BH property corrections  are included for parameterized tests as discussed in Sec.~\ref{sec:waveform}.
        Constraints for the GW150914- and GW170729-like events are assumed to be made with the third-generation detector CE for demonstration purposes, while the SMBH ones are assumed to have been observed with space-based detector LISA. Observe that additional corrections do not give significant contribution on bounding beyond-Kerr spacetimes with parameterized tests.
        }\label{tab:comparison}
\end{table}

Let us now investigate the effects of including ringdown and remnant BH corrections into the waveform.
In other words, how much does this change our results if only the inspiral corrections were included as is commonly done in parameterized tests?

We begin by performing parameterized tests in two separate cases: (i) with only ppE inspiral corrections present within the entire inspiral-merger-ringdown waveform, and (ii) with inspiral, ringdown, and remnant BH property corrections present in the waveform, as was done in the main analysis.
For demonstration purposes, we choose the third-generation detector CE observing GW150914-like events, GW170729-like events, and then space-based detector LISA observing SMBHB events as considered in the main analysis.
See Table~\ref{tab:comparison} for a summary of obtained results in each case.
We see that for the smaller-mass events GW150914 and GW170729, the two cases differ by up to $\sim15$\%.
As expected, the large-mass SMBH event observed by LISA only differs by $\sim3$\% due to the low-frequency window available to space-based detectors, where the inspiral corrections make the largest difference.
We conclude that such additional corrections to the ringdown and remnant BH properties in the waveform do not have significant contribution on constraining beyond-Kerr spacetimes with parameterized tests.

In fact, a similar feature can be seen for the IMR consistency tests. Since the systematic error in the merger-ringdown portion is typically smaller than that of the inspiral, even if we do not include corrections to the ringdown and final BH's mass and spin, we would still find bounds that are comparable to those presented in Table~\ref{tab:results}. 
These findings give us supporting evidence that in many cases, the dominant contribution comes from the corrections to the inspiral, as considered e.g. in~\cite{Yunes_ModifiedPhysics}.

\subsection{Modified-$\Delta$ spacetime}\label{sec:IMRDMD}
Now we repeat the analysis performed in Sec.~\ref{sec:results} in the mod.-$\Delta$ spacetime.
Because the results here are very similar to those found in the preceding section, we only outline a brief overview here.

\subsubsection{IMR consistency tests}
We begin by performing the IMR consistency test on GW150914-like and GW170729-like events observed on both CE, as well as with the multi-band observation between CE and LISA.
Figure~\ref{fig:imrKB} presents the resulting 90\% credible error ellipse in the $(M_f,\chi_f)$ plane for each case.
Similar to above in the JP spacetime, we observe that for O2, we can detect non-Kerr effects on GW150914-like events for $\beta\approx2$, and significantly higher at $\beta\approx14$  for GW170729-like events, due to the large inspiral uncertainty resulting from the large BH masses.
Such results are still less reliable than the following ones due to the large deviations, and we do not present the resulting contours, however the constraints are still tabulated in Tab.~\ref{tab:results}.
Following this, we see that when observed on future detector CE, GW150914-like events can resolve non-Kerr effects at a significantly smaller $\beta\approx0.05$, and a very similar value of $\beta\approx0.06$ for GW170729-like events.
Finally, we observe constraints of $\beta\approx5\times10^{-3}$ (GW150914-like) and $\beta\approx0.05$ (GW170729-like) for the multiband observations between CE and LISA.
The former strong constraint is a result of the small inspiral statistical uncertainties and large systematic uncertainties.

Following this, we repeat the IMR consistency test for LISA observations of EMRIs and SMBHBs.
Heeding the warning discussed previously in Sec.~\ref{sec:IMRJP} about the validity of EMRIs in the IMRPhenomD waveform, we present these results in Fig.~\ref{fig:imrKBmassive}.
Once again, the inspiral statistical uncertainty on EMRI observations is minuscule, resulting in the strong constraint of  $\beta\approx2\times10^{-4}$.
We observe inconsistencies between the inspiral and merger-ringdown signals in a SMBHB event at $\beta\approx10^{-3}$.
Finally, we note that in the mod.~$\Delta$ spacetime, typically the direction of systematic uncertainties in the $(M_f,\chi_f)$ plane are opposite to those in the JP spacetime. 
We found that this is primarily due to the different PN orders at which each spacetime alters the inspiral waveform at ($2$PN order in JP, $1$PN order in mod.~$\Delta$).
This effect is dominant among the corrections provided in this analysis, and serves to shift the direction of systematic uncertainties present in each spacetime.

\subsubsection{Parameterized tests}

In addition, we perform a set of parameterized tests of GR for each case considered here for comparison to the ones found with the IMR consistency test.
As in the JP case, we find that such bounds are comparable for each case discussed in this section.
As discussed in Sec.~\ref{sec:IMRJP}, the IMR consistency test is less valid due to the invalid use of the IMRPhenomD waveform, while such pieces were removed for the parameterized test.
We refer the reader to Table~\ref{tab:results} for a comprehensive display of all results found in this section.


\section{Conclusion and Discussion}\label{sec:conclusion}

Parameterized BH solutions to modified Einstein's field equations allow us to test the extreme-gravity regime of GR in a model-independent way.
One can obtain such a spacetime metric by parametrically deviating away from the famous Kerr spacetime metric with one or more parameters.
From here, parameterized corrections to the gravitational waveform for inspiraling BHs can be predicted.
Once one has these tools in hand, future GW signals can be tested against the beyond-Kerr metric by (i) IMR consistency tests (comparing the consistency between the inspiral and merger-ringdown portions of the signal) and (ii) parameterized tests.

In this paper, we presented the necessary recipe required to estimate corrections to the inspiral, ringdown, and remnant BH properties of the gravitational waveform, and then test future GW signals against this template with the above two tests.
In particular, we extended the work of Refs.~\cite{Carson_QNMPRD,Carson_QNM_PRL} by the same authors where this was done for the specific example of the Einstein-dilaton Gauss-Bonnet theory of gravity.
We first derive corrections in a generic way without specifying the beyond-Kerr spacetime. 
As an application, we focused on the JP metric introduced by Johannsen and Psaltis~\cite{Johannsen:2011dh} and the modified-$\Delta$ metric, modified from Johannsen's metric in Refs.~\cite{Johannsen:2015mdd,Johannsen:2015pca}.
Each spacetime metric considered here are singly-parameterized beyond the Kerr metric with parameters $\epsilon_3$ and $\beta$ respectively.
Such spacetimes can then be mapped to BH solutions other than Kerr found in the literature.

With the arbitrary JP and mod.-$\Delta$ metrics in hand, we next estimated corrections to the gravitational waveform for inspiraling BHs immersed in a JP or mod.-$\Delta$ spacetime.
Specifically, we found corrections to the GW amplitude and phase in the inspiral, the ringdown and damping QNM frequencies, the orbital energy and angular momentum of a particle about the BH, and finally the remnant BH's mass and spin.
Each of the above-listed corrections are parameterized by the single parameters $\epsilon_3$ and $\beta$ in the JP and mod.-$\Delta$ spacetimes respectively, and can be accordingly added into the gravitational waveform template.

We next performed the IMR consistency test to predict the magnitude of $\epsilon_3$ ($\beta$) required to differentiate between Kerr and JP (mod.-$\Delta$) GW signals.
Within this test, we first computed statistical uncertainties on the remnant BH mass and spin parameters from the inspiral and merger-ringdown signals independently, using a Fisher analysis.
We next estimated the systematic uncertainties in each measurement representing the waveform mismodeling uncertainty present by using a GR template with Kerr BHs, and yet observing a GW signal with a given magnitude of $\epsilon_3$ ($\beta$) present within.
We then increased the magnitude of $\epsilon_3$ ($\beta$) until the inspiral and merger-ringdown estimates of remnant BH properties begin to disagree to a statistically significant level.
Only at this point can we reliably claim the observed GW signal indeed has emergent  JP (mod.-$\Delta$) effects present within.
We also computed bounds on $\epsilon_3$ ($\beta$) using the parameterized test and compared them with those from the IMR consistency test.

We now discuss our findings.
We performed the IMR consistency test in each considered spacetime metric for the current-generation aLIGO O2 detector, the third-generation CE detector, the future space-based detector LISA, and finally the multi-band observation between the latter two.
As summarized in Table~\ref{tab:results}, we first found that observations by the O2 detector can detect JP (mod.-$\Delta$) deviations from the GR waveform for magnitudes of $\epsilon_3$ ($\beta$) on the order of unity, in agreement with current constraints.
For future GW detectors CE and LISA, we found that constraints about two orders-of-magnitude stronger can be claimed.
Finally, for the observation of EMRIs by the space-based detector LISA, we found that constraints three orders-of magnitude stronger can be made.
Such strong constraints occur because EMRI BH systems radiate GWs less compared to comparable-mass systems with the same total mass, thus increasing the amount of time JP (mod.-$\Delta$) effects are observed for, which results in a factor of $\eta^{-4/5}\sim10^{4}$ for EMRI systems in the ppE correction to the inspiral waveform.

In this analysis, several assumptions were made that somewhat weaken our results. 
In particular, we have assumed the following caveats:
\begin{itemize}
    \item We only included corrections to the ringdown phase of the waveform, neglecting those to the merger. 
    \item We only consider conservative corrections to the inspiral waveform, rather than dissipative ones. The resulting presented bounds are therefore conservative in nature. Once dissipative effects are additionally included, constraints may become stronger.
    \item We only included corrections to the leading-order PN terms in the waveform, and also to quadratic order in spin, and first order in beyond-Kerr parameters $\zeta$.
    \item We assumed that the QNMs are isospectral between axial and polar modes, something that may not be entirely true in beyond-Kerr spacetimes. 
    \item We estimated the BH final mass and spin following the result that holds for Kerr BH binary mergers in GR, which may not be true in beyond-Kerr spacetimes. 
\end{itemize}
One needs to specify a theory of gravity to overcome most of the points raised above, which goes beyond the scope of probing beyond-Kerr spacetime in a generic, model-independent way with GWs.
We present this article as a new method to quickly and easily estimate various corrections in the full waveform from an arbitrary beyond-Kerr metric, to obtain order-of-magnitude parameter constraints.
Future analyses could improve upon this work for more valid, yet significantly slower and computationally expensive results.
Specifically, repeated calculations with the more-robust Bayesian parameter-estimation analysis could be performed.
One could also study higher PN order-corrections and higher spin corrections beyond $\mathcal{O}(\epsilon_3,\beta,\chi^2)$ to the gravitational waveform\footnote{Higher-order spin contributions are estimated in Einstein-dilaton Gauss-Bonnet gravity and found to be negligible~\cite{Carson_QNMPRD}.}. 
Another avenue for future work includes studying beyond-Kerr spacetimes other than those considered here.


\section*{Acknowledgments}\label{acknowledgments}
Z.C. and K.Y. acknowledge support from NSF Award PHY-1806776. K.Y. would like to also acknowledge support by a Sloan Foundation Research Fellowship, the Ed Owens Fund, the COST Action GWverse CA16104 and JSPS KAKENHI Grants No. JP17H06358.


\appendix

\begin{widetext}
\section{Arbitrary remnant BH mass and spin corrections}\label{appendix}
In this appendix, we display the lengthy corrections to the remnant BH mass and spin given in an arbitrary spacetime metric $g_{\alpha\beta}^\X=g_{\alpha\beta}^\K+\zeta h_{\alpha\beta}^\X$ for general deviation parameter $\zeta$ and perturbation metric $h_{\alpha\beta}^\X$. The perturbation metric is further expanded up to quadratic order in BH spin as $h_{\alpha\beta}^\X=h_{\alpha\beta,0}+h_{\alpha\beta,1}\chi_f+h_{\alpha\beta,2}\chi_f^2$.
In the following expressions, all of the metric components are to be evaluated at $r=6M$.

\begin{gather}
\nonumber\delta M_f^\X=\frac{\mu}{23328 M_t^2}  \Bigg\lbrack2 \chi _f^\K \Big(62208 \sqrt{3} M_t^4 h_{\text{tt},0}''+15552 \sqrt{3} M_t^3 h_{\text{tt},0}'-23328 \sqrt{2} M_t^3 h_{\text{tt},1}'+10368 \sqrt{2} M_t^3 h_{\text{t$\phi $},0}''+432 \sqrt{3} M_t^2 h_{\text{tt},0}\\
\nonumber-3888 \sqrt{2} M_t^2 h_{\text{tt},1}-2376 \sqrt{2} M_t^2 h_{\text{t$\phi $},0}'-2592 \sqrt{3} M_t^2 h_{\text{t$\phi $},1}'+288 \sqrt{3} M_t^2 h_{\phi \phi ,0}''+432 \sqrt{2} M_t h_{\text{t$\phi $},0}+216 \sqrt{3} M_t h_{\text{t$\phi $},1}\\
\nonumber-204 \sqrt{3} M_t h_{\phi \phi ,0}'-108 \sqrt{2} M_t h_{\phi \phi ,1}'+55 \sqrt{3} h_{\phi \phi ,0}+36 \sqrt{2} h_{\phi \phi ,1}\Big)+\left(\chi _f^\K\right)^2 \Big(-248832 \sqrt{2} M_t^5 h_{\text{tt},0}'''-64800 \sqrt{2} M_t^4 h_{\text{tt},0}''\\
\nonumber+124416 \sqrt{3} M_t^4 h_{\text{tt},1}''-27648 \sqrt{3} M_t^4 h_{\text{t$\phi $},0}'''-1296 \sqrt{2} M_t^3 h_{\text{tt},0}'+31104 \sqrt{3} M_t^3 h_{\text{tt},1}'-46656 \sqrt{2} M_t^3 h_{\text{tt},2}'+12384 \sqrt{3} M_t^3 h_{\text{t$\phi $},0}''\\
\nonumber+20736 \sqrt{2} M_t^3 h_{\text{t$\phi $},1}''-1152 \sqrt{2} M_t^3 h_{\phi \phi ,0}'''+1332 \sqrt{2} M_t^2 h_{\text{tt},0}+864 \sqrt{3} M_t^2 h_{\text{tt},1}-7776 \sqrt{2} M_t^2 h_{\text{tt},2}-4344 \sqrt{3} M_t^2 h_{\text{t$\phi $},0}'\\
\nonumber-4752 \sqrt{2} M_t^2 h_{\text{t$\phi $},1}'-5184 \sqrt{3} M_t^2 h_{\text{t$\phi $},2}'+1332 \sqrt{2} M_t^2 h_{\phi \phi ,0}''+576 \sqrt{3} M_t^2 h_{\phi \phi ,1}''+880 \sqrt{3} M_t h_{\text{t$\phi $},0}+864 \sqrt{2} M_t h_{\text{t$\phi $},1}\\
\nonumber+432 \sqrt{3} M_t h_{\text{t$\phi $},2}-741 \sqrt{2} M_t h_{\phi \phi ,0}'-408 \sqrt{3} M_t h_{\phi \phi ,1}'-216 \sqrt{2} M_t h_{\phi \phi ,2}'+183 \sqrt{2} h_{\phi \phi ,0}+110 \sqrt{3} h_{\phi \phi ,1}+72 \sqrt{2} h_{\phi \phi ,2}\Big)\\
\nonumber-72 \sqrt{2} \Big(648 M_t^3 h_{\text{tt},0}'+108 M_t^2 h_{\text{tt},0}+36 \sqrt{6} M_t^2 h_{\text{t$\phi $},0}'-3 \sqrt{6} M_t h_{\text{t$\phi $},0}+3 M_t h_{\phi \phi ,0}'-h_{\phi \phi ,0}\Big)
-720 \sqrt{2} M_t^2 \chi _f^\K \delta \chi _f^\X-432 \sqrt{3} M_t^2 \delta \chi _f^\X\Bigg\rbrack\\
\end{gather}

\begin{gather}
\nonumber\delta\chi_f^\X=\frac{1}{6912}  \Bigg\{746496 \sqrt{6} h_{\text{tt},0}'' M_t^2+171072 \sqrt{6} h_{\text{tt},0}' M_t-559872 h_{\text{tt},1}' M_t+248832 h_{\text{t$\varphi $},0}'' M_t+2592 \sqrt{6} h_{\text{tt},0}+46656 h_{\text{tt},1}\\
\nonumber+\frac{648 \sqrt{6} h_{\varphi \varphi ,0}}{M_t^2}+\frac{1512 h_{\varphi \varphi ,1}}{M_t^2}-62208 h_{\text{t$\varphi $},0}'-31104 \sqrt{6} h_{\text{t$\varphi $},1}'+3456 \sqrt{6} h_{\varphi \varphi ,0}''+\frac{1}{M_t^2 \mu }\Bigg\lbrack\sqrt{3} \left(-9 M_t-6 \sqrt{2} \mu +\kappa  \sqrt{3}\right)\\
\nonumber \times\Big(-746496 h_{\text{tt},0}''' M_t^5-349920 h_{\text{tt},0}'' M_t^4+186624 \sqrt{6} h_{\text{tt},1}'' M_t^4-41472 \sqrt{6} h_{\text{t$\varphi $},0}''' M_t^4-42768 h_{\text{tt},0}' M_t^3+42768 \sqrt{6} h_{\text{tt},1}' M_t^3\\
\nonumber-139968 h_{\text{tt},2}' M_t^3+9936 \sqrt{6} h_{\text{t$\varphi $},0}'' M_t^3+62208 h_{\text{t$\varphi $},1}'' M_t^3-3456 h_{\varphi \varphi ,0}''' M_t^3+2106 h_{\text{tt},0} M_t^2+648 \sqrt{6} h_{\text{tt},1} M_t^2+11664 h_{\text{tt},2} M_t^2\\
\nonumber-4320 \sqrt{6} h_{\text{t$\varphi $},0}' M_t^2-15552 h_{\text{t$\varphi $},1}' M_t^2-7776 \sqrt{6} h_{\text{t$\varphi $},2}' M_t^2+3276 h_{\varphi \varphi ,0}'' M_t^2+864 \sqrt{6} h_{\varphi \varphi ,1}'' M_t^2+852 \sqrt{6} h_{\text{t$\varphi $},0} M_t+2376 h_{\text{t$\varphi $},1} M_t\\
\nonumber+2592 \sqrt{6} h_{\text{t$\varphi $},2} M_t-1677 h_{\varphi \varphi ,0}' M_t-630 \sqrt{6} h_{\varphi \varphi ,1}' M_t-648 h_{\varphi \varphi ,2}' M_t+389 h_{\varphi \varphi ,0}+162 \sqrt{6} h_{\varphi \varphi ,1}+378 h_{\varphi \varphi ,2}\Big)\Bigg\rbrack\\
\nonumber-\frac{12}{\kappa M_t^2} \Bigg\lbrack 48 \mu  \left(2592 h_{\text{tt},0}' M_t^3-216 h_{\text{tt},0} M_t^2+144 \sqrt{6} h_{\text{t$\varphi $},0}' M_t^2-48 \sqrt{6} h_{\text{t$\varphi $},0} M_t+12 h_{\varphi \varphi ,0}' M_t-7 h_{\varphi \varphi ,0}\right)\\
\nonumber+18 \left(3 M_t+2 \mu  \sqrt{2}\right) \Big(10368 \sqrt{2} h_{\text{tt},0}'' M_t^4+2376 \sqrt{2} h_{\text{tt},0}' M_t^3-2592 \sqrt{3} h_{\text{tt},1}' M_t^3+1152 \sqrt{3} h_{\text{t$\varphi $},0}'' M_t^3+36 \sqrt{2} h_{\text{tt},0} M_t^2\\
\nonumber+216 \sqrt{3} h_{\text{tt},1} M_t^2-288 \sqrt{3} h_{\text{t$\varphi $},0}' M_t^2-432 \sqrt{2} h_{\text{t$\varphi $},1}' M_t^2+48 \sqrt{2} h_{\varphi \varphi ,0}'' M_t^2+44 \sqrt{3} h_{\text{t$\varphi $},0} M_t+144 \sqrt{2} h_{\text{t$\varphi $},1} M_t-35 \sqrt{2} h_{\varphi \varphi ,0}' M_t\\
\nonumber-12 \sqrt{3} h_{\varphi \varphi ,1}' M_t+9 \sqrt{2} h_{\varphi \varphi ,0}+7 \sqrt{3} h_{\varphi \varphi ,1}\Big)+\left(4 \mu  \sqrt{3}+(M_t \delta_m+\mu\lambda ) \chi _a+(M_t+ \delta_m \mu\lambda ) \chi _s\right) \Big(-746496 \sqrt{3} h_{\text{tt},0}''' M_t^5\\
\nonumber-349920 \sqrt{3} h_{\text{tt},0}'' M_t^4+559872 \sqrt{2} h_{\text{tt},1}'' M_t^4-124416 \sqrt{2} h_{\text{t$\varphi $},0}''' M_t^4-42768 \sqrt{3} h_{\text{tt},0}' M_t^3+128304 \sqrt{2} h_{\text{tt},1}' M_t^3\\
\nonumber-139968 \sqrt{3} h_{\text{tt},2}' M_t^3+29808 \sqrt{2} h_{\text{t$\varphi $},0}'' M_t^3+62208 \sqrt{3} h_{\text{t$\varphi $},1}'' M_t^3-3456 \sqrt{3} h_{\varphi \varphi ,0}''' M_t^3+2106 \sqrt{3} h_{\text{tt},0} M_t^2+1944 \sqrt{2} h_{\text{tt},1} M_t^2\\
\nonumber+11664 \sqrt{3} h_{\text{tt},2} M_t^2-12960 \sqrt{2} h_{\text{t$\varphi $},0}' M_t^2-15552 \sqrt{3} h_{\text{t$\varphi $},1}' M_t^2-23328 \sqrt{2} h_{\text{t$\varphi $},2}' M_t^2+3276 \sqrt{3} h_{\varphi \varphi ,0}'' M_t^2+2592 \sqrt{2} h_{\varphi \varphi ,1}'' M_t^2\\
\nonumber+2556 \sqrt{2} h_{\text{t$\varphi $},0} M_t+2376 \sqrt{3} h_{\text{t$\varphi $},1} M_t+7776 \sqrt{2} h_{\text{t$\varphi $},2} M_t-1677 \sqrt{3} h_{\varphi \varphi ,0}' M_t-1890 \sqrt{2} h_{\varphi \varphi ,1}' M_t-648 \sqrt{3} h_{\varphi \varphi ,2}' M_t+389 \sqrt{3} h_{\varphi \varphi ,0}\\
+486 \sqrt{2} h_{\varphi \varphi ,1}+378 \sqrt{3} h_{\varphi \varphi ,2}\Big)\Bigg\rbrack+\frac{1}{M_t}\left(9504 h_{\text{t$\varphi $},0}+10368 \sqrt{6} h_{\text{t$\varphi $},1}-2520 \sqrt{6} h_{\varphi \varphi ,0}'-2592 h_{\varphi \varphi ,1}'\right)\Bigg\}
\end{gather}
\end{widetext}


\bibliography{Zack}
\end{document}

%% file: preamble.tex
\usepackage{mathtools}
\usepackage{graphics}
\usepackage{graphicx}
\usepackage{dcolumn}
\usepackage{bm}
\usepackage{dsfont} 
\usepackage{amsmath,amssymb}
\usepackage{verbatimbox}
\usepackage{hyperref}
\usepackage{tabularx}
\usepackage{epsf,epsfig}
\usepackage[normalem]{ulem}
\usepackage[usenames]{color}
\usepackage{multirow}
\usepackage{makecell}
\usepackage{diagbox}
\usepackage[abs]{overpic}
\allowdisplaybreaks
\hypersetup{
    colorlinks=true,
    linkcolor=blue,
    filecolor=magenta,      
    urlcolor=blue,
    citecolor=blue
}
\newcommand{\R}{{\mbox{\tiny R}}}
\newcommand{\I}{{\mbox{\tiny I}}}
\newcommand{\ISCO}{{\mbox{\tiny ISCO}}}

\newcommand{\ppE}{{\mbox{\tiny ppE}}}
\newcommand{\GR}{{\mbox{\tiny GR}}}
\newcommand{\X}{{\mbox{\tiny X}}}

\newcommand{\MR}{{\mbox{\tiny MR}}}
\newcommand{\GW}{{\mbox{\tiny GW}}}

\newcommand{\K}{{\mbox{\tiny K}}}
\newcommand{\A}{{\mbox{\tiny A}}}
\newcommand{\JP}{{\mbox{\tiny JP}}}
\newcommand{\MD}{{\mbox{\tiny MD}}}

\definecolor{red(ncs)}{rgb}{0.77, 0.01, 0.2}

\usepackage{array}
\newcolumntype{C}[1]{>{\centering\let\newline\\\arraybackslash\hspace{0pt}}m{#1}}

\newcolumntype{C}[1]{>{\centering\arraybackslash}m{#1}}